\documentclass[10pt,a4paper,pointlessnumbers]{scrartcl}

\usepackage{lscape,epsfig,mytools}
\usepackage[hang,small,sf,up]{caption}

\def\PDG{{\sc Particle Data Group~}}

\def\Id{{\rm 1\kern-.3em I}}
\def\PDG{{\sc Particle Data Group~}}
\def\ie{{\it i.e.~}}
\def\eg{{\it e.g.~}}
\def\etal{{\it et al.~}}

\def\B{{$\cal B$~}}
\def\TOP{{\:\textsf{T}\:}}
\def\tr{{\:\textsf{tr}\:}}

\def\123{{\cal M}_1\to{\cal M}_2{\cal M}_3}

\newcommand{\equ}[1]{ \begin{eqnarray}  #1 \end{eqnarray}}

\newcommand{\myslash}[1]{{ #1 \!\!\!\!\!\!\:\: /} }

\begin{document}

\title{Strong Two--Body Decays of Light Mesons}

\author{\sf Ralf Ricken, Matthias Koll\thanks{e-mail adress: {\tt
koll@itkp.uni-bonn.de}} , Dirk Merten, Bernard C. Metsch} 

\date{\small\sf --- {\today} ---}

\maketitle

\begin{abstract}
In this paper, we present results on strong two-body decay widths of
light $q\bar q$ mesons calculated in a covariant quark model. The
model is based on the Bethe-Salpeter equation in its instantaneous
approximation and has already been used for computing the complete
meson mass spectrum and many electroweak decay observables. Our
approach relies on the use of a phenomenological confinement potential
with an appropriate spinorial Dirac structure and 't~Hooft's
instanton--induced interaction as a residual force for pseudoscalar
and scalar mesons. \par The transition matrix element for the decay of
one initial meson into two final mesons is evaluated in lowest order
by considering conventional decays via quark loops as well as Zweig
rule violating instanton--induced decays generated by the six--quark
vertex of 't~Hooft's interaction; the latter mechanism only
contributes if all mesons in the decay have zero total angular
momentum. We show that the interference of both decay mechanisms plays
an important role in the description of the partial widths of scalar 
and pseudoscalar mesons. 
\end{abstract}


\section{Introduction}\label{sec:intro}

The study of the internal structure of mesons is still a challenge to theoretical
physics as well as to experimental physics. Despite all efforts undertaken in
the last decades, in particular the strong decays of $q\bar q$ bound
states are rather poorly understood. For theory, one of the reasons might be
related to the fact that --- at least for the sector of mesons being composed out
of light quarks --- a relativistic treatment of the underlying
dynamics seems to be mandatory. In principle, this requirement is
satisfied by a study of strong two--body decays on the basis of the
Bethe--Salpeter equation in its instantaneous approximation. However,
the particular structure of the full transition operator for this
class of mesonic decays is quite unclear even in a covariant framework.
In the present work, we suppose that the pure quark loop contribution
present in all strong decays is accompanied by an additional
instanton--induced decay mechanism if and only if all mesons in the process 
have a vanishing total angular momentum.

Before we discuss our approach to the problem of strong two--body
decays of light mesons in detail, let us briefly make some remarks on
a non--relativistic phenomenological model which is quite prominent in
this field, namely the so--called $^3P_0$ model. It has its origins in
the early works of A.~Le~Yaouanc \etal published in
refs. \cite{LeYaouanc1,LeYaouanc2,LeYaouanc3} (see also
\cite{LeYaouanc4}). As has been suggested 
earlier by L.~Micu (see \cite{Micu}), the authors assume that these
two--body strong decays proceed via the creation of a $q\bar q$ state
with vacuum quantum numbers $J^{\pi c}=0^{++}$; the equivalent
spectroscopical notation for $q\bar q$ bound states reads 
$^{2S+1}L_J={}^3P_0$ which has in fact labeled this particular model. In ref. 
\cite{BarnesStrongDecays1}, T.~Barnes \etal have reviewed the results
of the modern $^3P_0$ model with respect to strong two--body
decays and compared the results to the
(rare and ambiguous) experimental data for light $n\bar n$ ground--state mesons;
higher quarkonia are studied in ref.  \cite{BarnesStrongDecays2}. For
completeness, we should also mention the recently published work of
R.~Bonnaz and B.~Silvestre-Brac  who include instanton induced effects
as well as tensor forces
in their study on the basis of the $^3P_0$ model (see
\cite{BonnazSilvestreBrac} and references therein); for decays of
non--$n\bar n$ meson states, we will refer to their
results.\footnote{Note that a recent preprint by Barnes {\sl et al.}
(see \cite{Barnes:2002mu}) also investigates the non-$n\bar n$ meson decays; we comment on this
work in section 4.}

Our Bethe--Salpeter approach has first been discussed in
refs. \cite{ResagMünzFirstPaper,ResagMünzSecondPaper}; recently, we
have presented an updated review concerning meson spectra and
electroweak decay characteristics (see
\cite{KollRicken,RickenKoll}). In the present paper, we aim at a
rather complete description of 
strong two--body decays of light mesons in the framework of our
relativistic quark model. Our approach is furthermore interesting
insofar as we not only compute the quark loop contribution but also an
instanton--induced term for $J_i=0$ mesons as well as the
interference of both mechanism; note that the latter interaction has already been
studied separately in ref. \cite{RitterPaper} by Ch.~Ritter \etal for
the present relativistic framework. 

For the following comprehensive discussion, we compare our results
with the availiable  experimental data compiled by the \PDG (see
\cite{PDG2000}; additional results can be found in
\cite{Abele,Suh,Barberis,Bugg,Reinnarth}) on the one hand as
well as with the results of the $^3P_0$ model presented in
refs.
\cite{BarnesStrongDecays1,BarnesStrongDecays2,BonnazSilvestreBrac} on
the other hand. Thus, we 
understand the current work as a reference frame for a more reliable
assessment of the multifaceted (and partly contradicting) results on
strong two--body decays of light mesons.

We have organized this contribution as follows: in section
\ref{BSEIntro}, we briefly review our covariant approach on the basis
of the Bethe--Salpeter equation in its instantaneous approximation;
furthermore, we comment on the results concerning the light meson
spectra (see also \cite{KollRicken,RickenKoll}). The transition matrix
element for strong decays of the type $\123$ is derived in section
\ref{app:strongtrans}; as mentioned above, it includes an
instanton--induced contribution for scalar and pseudoscalar mesons
beyond the lowest order quark loop part. Our results for numerous
decays are presented and discussed in section \ref{Results}. Finally,
we give a summary of our work in section \ref{Summary}. The appendix
includes technical details concerning the flavour matrix elements and
G.~'t~Hooft's instanton--induced interaction following
refs. \cite{tHooft,ShifmanVainshteinZakharov,PetryInst}.


\section{A Relativistic Quark Model for Mesons}
\label{BSEIntro}

In our approach, we treat the (constituent) quarks as fundamental degrees of freedom
for the description of hadronic bound states. The resulting quark
model is formally covariant and relies basically on the instantaneous
Bethe--Salpeter equation. It has been reviewed in detail in
refs. \cite{KollRicken,RickenKoll} such that we can keep the following
introduction rather compact before we turn to the strong decay matrix
elements in the next section.


\subsection{The instantaneous Bethe-Salpeter approach}\label{sect:BSapproach}

In quantum field theory, a quark--anti-quark bound state with
four--momentum $P$ and mass $M$ ($M^2 = P^2$) is described by the
Bethe--Salpeter equation for two fermions (see \cite{BetheSalpeter}). In
momentum space, this equation reads
\begin{eqnarray}
  \label{BS-Gleichung}
  \chi^{P} (p) &=& -\:i \:\: S_1^F (\frac{P}{2}+p)  \left[
  \int \frac{d^4 p'}{(2\pi)^4}  K(P, p, p') \chi^{P}
  (p')\right] S_2^F(-\frac{P}{2}+p)
\end{eqnarray}
where $p$ is the relative four--momentum between the quark and the
anti-quark, $K$ denotes the infinite sum of their irreducible
interactions and the corresponding full Feynman propagators are
labeled by $S^F_1$ and $S^F_2$, respectively. The Bethe--Salpeter
amplitude $\chi^P$ is defined in coordinate space as the time--ordered
product of the quark and the anti-quark field operator between the
bound state $|P\rangle$ and the vacuum: 
\begin{eqnarray}        
\label{BSA}
   \chi ^{P}_{\alpha\beta} (x_1, x_2) &:=&  
   \left\langle\: 0 \: \left |
       \TOP \:\psi_\alpha^1 (x_1)\bar\psi^2_\beta (x_2)\right |\:P
     \:\right\rangle\nonumber\\ 
&=& e^{-iP\cdot(x_1 + x_2)/2} \int\frac{d^4 p}{(2\pi)^4}
  e^{-ip\cdot(x_1 - x_2)} \chi ^{P}_{\alpha\beta} (p)
\end{eqnarray}
where $\alpha$ and $\beta$ are multi--indices for the Dirac, flavour and colour
degrees of freedom which are omitted in the following
discussion. Since in general the interaction kernel $K$ and the full
quark propagators 
$S^F_i$ are unknown quantities, we make two (formally covariant) approximations:
\begin{itemize}
\item The propagators are assumed to be of the free form $S^F_i = i
  (\;\myslash p - m_i + i\epsilon\;)^{-1}$ with effective constituent quark masses
  $m_i$ that we consider as free parameters in our model; thus, all
dynamical self--interactions of the (anti-)quarks are believed to be
adequately parameterized by a constant absorbed in the constituent masses.
\item The interaction kernel shall only depend on the components
  of $p$ and $p\pri$ perpendicular to $P$, \ie we assume that $K(P,p,p\pri) = V(p_{\perp P},
  p\pri_{\perp P})$ with $p_{\|P}:=(p\cdot P/P^2)P$ and $p_{\perp P} := p
- p_{\|P}$ holds; therefore, the interaction kernel in the meson rest
frame with $P_M=(M,\vec 0)$ becomes independent of the 
variable $p_{\|P_M}\equiv p^0$ which is the reason for labelling this
assumption the ``instantaneous approximation''.
\end{itemize}

Integrating in the bound state rest frame over the relative energy
$p^0$ and
introducing the equal--time  amplitude (or Salpeter amplitude)
$\Phi(\vec p\;):=\int\frac{dp^0}{2\pi}\chi^P (p^0, \vec p\;)|_{P=(M, \vec
0)}$, we end up with the Salpeter equation (see \cite{Salpeter}) which constitutes
the basic equation of our model:
\begin{eqnarray}\label{eq:SE}
  \Phi (\vec p\;) &=& + \:\Lambda_1^-(\vec p\;) \gamma ^0 \left[ 
    \int\frac{d^3 p'}{(2\pi)^3} \frac{V(\vec p, \vec p\, ') 
      \Phi (\vec p\, ')}{M+\omega _1 + \omega _2} \right] 
  \gamma ^0\Lambda _2^+ (-\vec p\;)\nonumber \\
  & & -\: \Lambda_1^+(\vec p\;) \gamma ^0 \left[ \int\frac{d^3
      p'}{(2\pi)^3}\frac{ V(\vec p, \vec p\, ') \Phi (\vec p\, ')}{M-\omega _1 -
      \omega _2} \right] \gamma ^0\Lambda _2^-  (-\vec p\;)\; .
\end{eqnarray}
Here, $\Lambda _i^\pm(\vec p\;) = \frac 1 2  \pm \gamma^0(\vec \gamma \cdot
\vec p + m_i)/2\omega_i$ are projectors on positive and negative energy
solutions of the Dirac equation and $\omega_i=\sqrt{\vec p\,^2 +
m_i^2}$ denotes the kinetic energy of the quarks. The simultaneous
calculation of the meson masses $M$ and the Salpeter amplitudes $\Phi$
results by solving the corresponding eigenvalue problem of
eq. (\ref{eq:SE}) with an adequate potential ansatz (see
\cite{ResagMünzFirstPaper,KollRicken} for further details). 

The calculation of transition matrix
elements for decays or scattering processes can be done in the
Mandelstam formalism (see \cite{Mandelstam}). To this end, the
full Bethe-Salpeter amplitude $\chi^P(p)$ depending on the relative
four--momentum $p$ has to be known. On the mass shell of the bound
state, it can be reconstructed from the Salpeter amplitude $\Phi(\vec
p\,)$ in a covariant manner. Let us therefore first define the
meson-quark-antiquark vertex function  $\Gamma^P(p) :=
[S^F_1(P/2+p)]^{-1}\chi^P(p)[S^F_2(-P/2+p)]^{-1}$ as the amputated
Bethe--Salpeter amplitude. Starting with the corresponding amputated Bethe--Salpeter
equation, the vertex function in the meson's rest frame  
can be computed from the three--dimensional Salpeter amplitude by
\begin{eqnarray}\label{eq:vertexrec}
\Gamma(\vec p\;) :=  \Gamma^P(p)|_{P=(M,\vec 0)}
  = - i \int \frac{d^3p\pri}{(2\pi)^3}
  V(\vec p, \vec p\;\pri)\Phi(\vec p\;\pri)\;.
\end{eqnarray}
Due to covariance of this procedure, we can finally calculate the Bethe--Salpeter
amplitude for any on--shell momentum $P'=(P_0',\vec P\,')$ of the bound
state by performing a pure boost $\Lambda$ according to 
\begin{eqnarray}
  \chi^{P'}(p) &=& S_{\Lambda}\;\; \chi^P
  (\Lambda^{-1}p)\;\; S^{-1}_{\Lambda} \quad\mbox{with}\quad  \chi^P
  (p) = \chi^P
  (p_0,\vec p\,) = S^F_1(\frac M 2 +p_0,\vec p\,)\:\Gamma(\vec p\,)\:S^F_2(-\frac M 2
  +p_0,\vec p\,)
\end{eqnarray}
where $S_{\Lambda}$ denotes the corresponding transformation in
Dirac space and the boost is defined by $\Lambda P=P'$ for a total
momentum $P=(M,\vec 0)$ in the rest frame of the bound state. 


\subsection{Interactions, Parameters and Spectra}
\label{IntParSpec}

Up to now, we have not specified the interactions that we will include
in the instantaneous interaction kernel. From phenomenological studies
of QCD, one finds that a confining interaction is mandatory for a
satisfying description of low--energy hadrons and especially of their
radially excited states. Furthermore, it is clear that an additional
interaction should be included in the discussion as a
flavour--dependent force is required at least in the sector with total
angular momentum $J=0$. Our candidate for this residual interaction
will be 't~Hooft's instanton--induced force (see
appendix \ref{tHooft2Body}). Let us briefly comment on both types of interactions:

\begin{itemize}

\item The confining interaction in coordinate space is mostly parameterized
as a  linearly rising potential of the form 
${\cal V}_C\left(r \right) = a_c +b_c\cdot r$. Moreover, one also has
to choose a particular spinorial structure ${\mit\Gamma}\otimes{\mit\Gamma}$ for
the confining interaction that acts in Dirac space and is {\sl a
priori} unknown. 

Accordingly, the confining interaction in momentum space acting on the
Salpeter amplitude can be written as

\equ{
\int\frac{d^3p'}{(2\pi)^3}\:V_C\left(\vec p, \vec p\:'\right)\Phi (\vec p\:') = 
\int\frac{d^3p'}{(2\pi)^3}\:\tilde{\cal V}_C\left((\vec p - \vec
p\:')^2\right)\:\cdot\:{\mit\Gamma}\Phi (\vec p\:'){\mit \Gamma}
}

where $\tilde {\cal V}_C((\vec p - \vec p\:')^2)$ is the Fourier transform of
the linear potential ${\cal V}_C(r)$ in coordinate space. Note that
the offset $a_c$ and the slope $b_c$ will be treated as free parameters in our model. 

\item As it has been shown in the framework of a non--relativistic quark
model (see \cite{NonRelITKPModel,MetschHabil}), the
instanton--induced interaction on the basis of the ideas of 't~Hooft
and others (see \cite{tHooft,ShifmanVainshteinZakharov,PetryInst}) provides a
remarkably good explanation for the ground state masses in the
pseudoscalar sector; in a relativistic formulation, 't~Hooft's force
also acts in the scalar meson sector. 

In momentum space, the
effective potential for the instanton--induced interaction  ({\sl
abbr.:} III) can be written as

\equ{
\label{tHooftKernelOnPhi}
{\int\frac{d^3p'}{(2\pi)^3}\:V_{\mbox{\scriptsize III}}\left(\vec
p, \vec p\:'\right)\Phi (\vec p\:')} =
4G(g,g')\int\frac{d^3p'}{(2\pi)^3}\:{\cal R}_\Lambda\left(\vec p,\vec
p\:'\right)\:\cdot\:\left(\Id\tr[\Phi (\vec p\:')] + \gamma _5
\tr[\gamma _5\Phi (\vec p\:')]\right) 
}

where $\Lambda=\Lambda_{\mbox{\scriptsize
III}}$ denotes the finite effective range of the
force and the regulator function is of Gaussian type, \ie it reads
${\cal R}_\Lambda (\vec x) =  \exp({-{|\vec x|^2}/{\Lambda^2}}) / (\Lambda\sqrt\pi)^3$ in configuration space.
The matrix $G(g,g')$ includes flavour dependent coupling constants $g$
and $g'$; a summation over flavour indices is implicitly understood (see
appendix \ref{tHooft2Body} for more details). We will consider the effective range
and the coupling constants of 't~Hooft's force as free parameters. 
\end{itemize}

We have shown in refs. \cite{KollRicken,RickenKoll} that these
assumptions on the underlying interactions between the constituents of
the mesons produce a consistent picture of the complete light meson
sector.\footnote{Note that a similar model for baryons described a
$qqq$ states has been presented in
refs. \cite{Löring1,Löring2,Löring3} by U.~L\"oring \etal; the authors also use the
instantaneous Bethe--Salpeter equation and adopt a confining force plus a
residual instanton--induced interaction.} The parameters that we have
introduced in these former publications were summarized in two sets denoted by
``$\cal A$'' and ``$\cal B$'', respectively. In this work, we will
only refer to a unique parameter set for the sake of clarity, namely
model $\cal B$; the specific numerical values for the corresponding
parameters are displayed in tab.~\ref{tab:Parameters}. At this point, let us stress that we do not
alter these parameters which are uniquely fixed to the meson mass spectrum
(see \cite{KollRicken,RickenKoll}). Thus, the following results on
strong decay widths can be considered as (quasi) parameter--free
predictions insofar as only the three--body 't~Hooft coupling
$g^{(3)}_{\mbox{\scriptsize eff}}$ has to
be fixed to a selected decay with $J_i=0$ ($i=1,2,3$); this particular
coupling constant appears here for the first time in our
model (see also section \ref{Fixing_g3}). However, we do not aim at a high--precision prediction of
strong decay widths with our approach; instead, we consider this work
merely as a global overview including qualitative as well as
quantitative features and thus providing a reliable framework for
future efforts in this field. 

In refs. \cite{KollRicken,RickenKoll}, we have thoroughly
discussed our results on the mass spectra in the light meson
sector. Concerning the ground states, we obtained an excellent
description of the well--known Regge trajectories up to highest
angular momenta. The combination of the particular spinorial structure
of the confinement force in our model and the effects of 't~Hooft's
instanton--induced interaction provides very good results for the
pseudoscalar mass spectrum; concerning the scalar sector, we observe a
plausible classification of $q\bar q$ mesons even for the first
excited states (see \cite{RickenKoll} for details). In general, the
ground states as well as the radial excitations in the isovector,
isodublet and isoscalar sectors are reasonably well described with the
parameters given in tab.~\ref{tab:Parameters} such that we consider
our calculation of the complete light meson mass spectrum as a good
starting point for the study of the strong decays of these bound
states.


\section{The Transition Matrix Element}
\label{app:strongtrans}

For the description of the decay of an initial meson  with
four--momentum $P_1$ into two final mesons with
four­-momenta $P_2$ and $P_3$, respectively, we consider the
transition matrix element  

\begin{eqnarray}
T_{P_1\rightarrow P_2 P_3}\; = \; \:\big\langle \:P_2 \:P_3\:\big|\: T
\:\big|\: P_1\:\big\rangle \quad.
\end{eqnarray}

As we will show in the following, the so--called Mandelstam formalism
(see \cite{Mandelstam}) allows for the calculation of any dynamical
bound state observable from the corresponding Bethe-Salpeter
amplitudes.  


\subsection{The Six--Point Green's Function}

The decay of one initial meson into two final mesons proceeds via an
interaction of six (anti--)quarks. Let us therefore define the
corresponding six--point Green's function in coordinate space according to
\begin{eqnarray}
\label{eq:G6}
G^{(6)}_{\alpha\alpha',\beta\beta',\gamma\gamma'}(x_1, x_2,
y_1, y_2, z_1, z_2) :=
\big\langle\:0\:\big|\TOP\Psi_{\alpha'}(y_1)\bar\Psi_{\beta'}(y_2)
\Psi_{\gamma'}(z_2)\bar\Psi_{\gamma}(z_1)\Psi_{\beta}(x_2)\bar\Psi_{\alpha}(x_1)\:\big|\:0\:\big\rangle\quad.   
\end{eqnarray}
For the following discussion, it is helpful to
introduce furthermore the four-point Green's function
\begin{eqnarray}
\label{eq:G4}
G^{(4)}_{\alpha_1\alpha, \beta_1\beta}(x_1', x_2', x_1, x_2):=
-\big\langle
\:0\:\big|\TOP\Psi_{\alpha_1}(x_1')\bar\Psi_{\beta_1}(x_2')\Psi_{\beta}(x_2)\bar\Psi_{\alpha}(x_1)\:\big|\:0
\:\big\rangle\;
\end{eqnarray}
as well as the eight-point Green's function given by
\begin{eqnarray}
\label{eq:G8}
\lefteqn{
G^{(8)}_{\alpha'\alpha_1',
\beta'\beta_1',\gamma\gamma_1,\gamma'\gamma_1'}(y_1, y_2, z_1, z_2,
y_1', y_2', z_1', z_2')  } \\ \nonumber &:=& - \big\langle\: 
0\:\big|\TOP\Psi_{\alpha'}(y_1)\bar\Psi_{\gamma}(z_1)\Psi_{\gamma_1}(z_1')
\bar\Psi_{\alpha_1'}(y_1')\Psi_{\gamma'}(z_2)\bar\Psi_{\beta'}(y_2)
\Psi_{\beta_1'}(y_2')\bar\Psi_{\gamma_1'}(z_2')\:\big|\:0 
\:\big\rangle\;. 
\end{eqnarray}

Here, the full Heisenberg field operators $\Psi$ and $\bar \Psi$ are labeled
with multi--indices $\alpha, \beta, \ldots$ refering to Dirac, flavour
and colour space. In these definitions of the $n$--point Green's functions $G^{(n)}$, the
symbol $\TOP$ denotes the time--ordering operator.

\begin{figure}[b!]
\begin{center}
\input{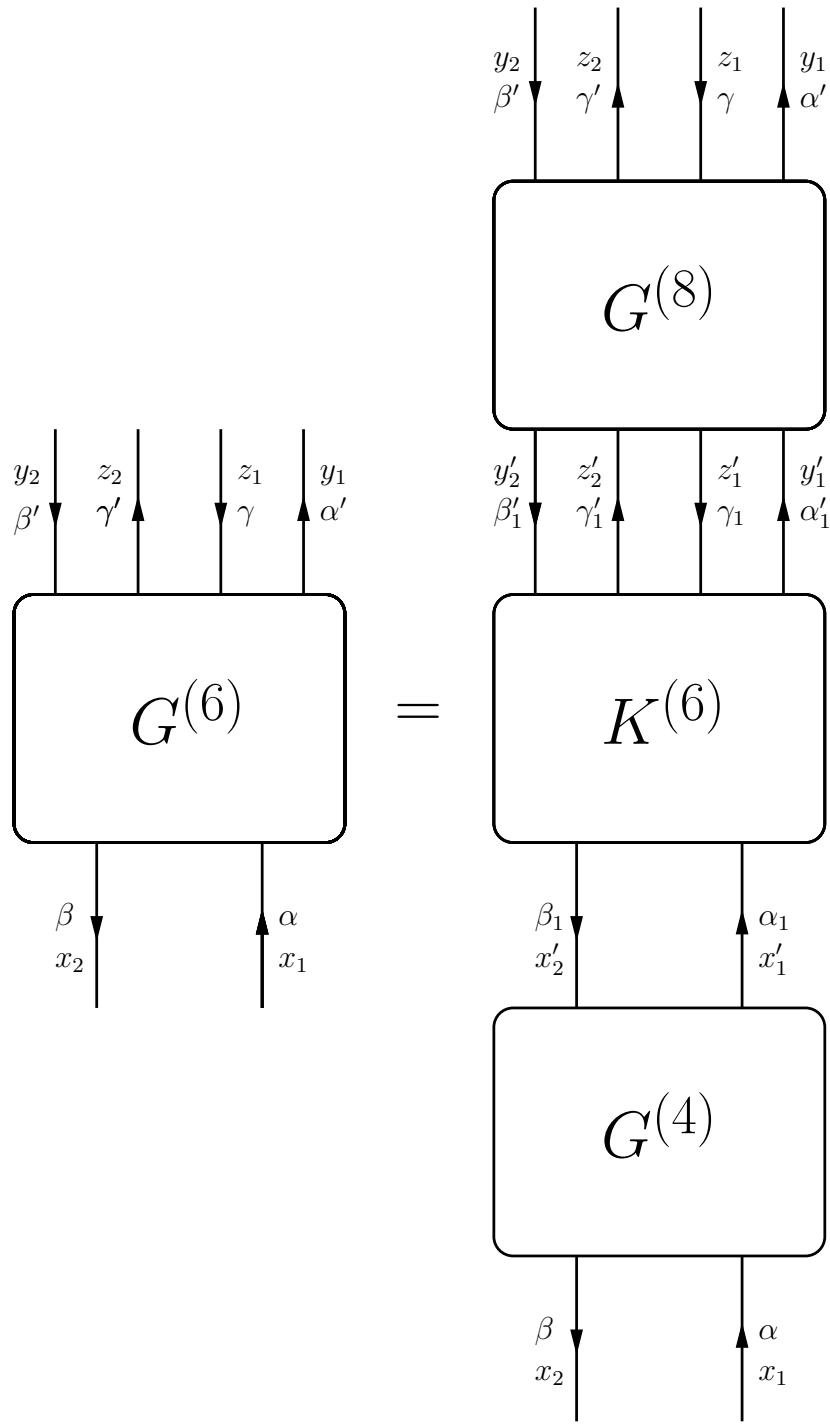}
\caption{Decomposition of the six--point Green's function $G^{(6)}$ into
  the corresponding irreducible kernel $K^{(6)}$.}
\label{fig:FULLKSIX}
\end{center}
\end{figure}

In order to connect the full six--point Green's function with the
bound state Bethe-Salpeter amplitudes of the three mesons, we first
decompose $G^{(6)}$ into a kernel $K^{(6)}$ which is irreducible with
respect to the four-point Green's function $G^{(4)}$ of the incoming
$\bar q q$ pair and the eight-point Green's function $G^{(8)}$ of the
two outgoing $\bar q q$ pairs: 
\begin{eqnarray}\label{eq:K6} 
G^{(6)}_{\alpha\alpha',\beta\beta',\gamma\gamma'}(x_1, x_2,
y_1, y_2, z_1, z_2) \!\!\!&=&\!\!\! \int
d^4x_1'd^4x_2'd^4y_1'd^4y_2'd^4z_1'd^4z_2' \:\:  G^{(8)}_{\alpha'\alpha_1',
\beta'\beta_1',\gamma\gamma_1,\gamma'\gamma_1'}(y_1, y_2, z_1, z_2,
y_1', y_2', z_1', z_2')\nonumber\\ 
&&\quad\times\quad K^{(6)}_{\alpha_1\alpha_1', \beta_1\beta_1',
\gamma_1\gamma_1'}(x_1', x_2', y_1', y_2', z_1', z_2')\:\:
G^{(4)}_{\alpha_1\alpha, \beta_1\beta}(x_1', x_2', x_1, x_2)\;. 
\end{eqnarray}
This equation defines the six--point kernel $K^{(6)}$; in
fig.~\ref{fig:FULLKSIX}, we give a diagrammatical representation of
this decomposition.

The four-point Green's function $G^{(4)}$ describes the propagation of
a $q\bar q$ pair within the space-time region where it is
created from and annihilated into the vacuum. As we want to calculate
the decay of a meson, we are interested in those contributions to
$G^{(4)}$ that originate from $q\bar q$ bound states denoted by $| P_i
\rangle$. Here and in the following, $P_i$ and $M_i$ denote the four--momentum and
the mass of the bound state ``$i$'' with $P_i^2=M^2_i$; the Fock states $| P_i
\rangle$ are normalized according to 
\begin{eqnarray}
\big\langle\: P_i\:\big|\:P_i'\:\big\rangle \; = \; (2\pi)^3\;
2\omega_{P_i}\; \delta(\vec P_i -\vec
P_i')\quad\mbox{with}\quad \omega_{P_i}\; := \; P_i^0 = \sqrt{M^2 +
\vec P_i^2} \:.
\end{eqnarray} 
Assuming the
time-ordering ${x_1}^0, {x_2}^0 < {{x_1'}^0}, {{x_2'}^0}$,  a complete
set of these states can be inserted into $G^{(4)}$ such that the
bound state contribution to the four--point Green's function is given by 
\begin{eqnarray}
\tilde G^{(4)}_{\alpha_1\alpha, \beta_1\beta}(x_1', x_2', x_1, x_2)
&:=& - \int d \tilde P_1 \:\: \big\langle\:  0\:\big| T \Psi_{\alpha_1}(x_1')\bar
\Psi_{\beta_1}(x_2')\:\big| P_1\: \big\rangle  \big\langle\: 
P_1\:\big|\TOP\Psi_{\beta}(x_2)\bar\Psi_{\alpha}(x_1)\:\big|0\: \big\rangle  \nonumber \\
&=& - \int d\tilde P_1 \:\:\chi^{P_1}_{\alpha_1\beta_1}(x_1',
x_2')\bar\chi^{P_1}_{\beta\alpha}(x_1, x_2)\; 
\end{eqnarray}
with the definition $d \tilde P := d^3P/ ((2\pi)^32\omega_P)$. Note
that we have used the definition of the Bethe-Salpeter amplitude
$\chi$ according to eq. (\ref{BSA}); an
analogous definition holds for the adjoint amplitude $\bar \chi$.
 
In order to express also the eight-point Green's
function $G^{(8)}$ by Bethe-Salpeter amplitudes alone, we make
the following fundamental approximation: we assume that the two
outgoing mesons do not interact with each other, \ie our ansatz
relies on the complete neglect of any final state interaction.
Then it is allowed to decompose the eight-point Green's
function into a product of two four-point Green's functions according
to $G^{(8)} \approx G^{(4)}\cdot G^{(4)} $. Considering again only the
bound state contributions to four-point Green's functions, we end
up with 
\begin{eqnarray}
\lefteqn{\tilde G^{(8)}_{\alpha'\alpha_1',
\beta'\beta_1',\gamma\gamma_1,\gamma'\gamma_1'}(y_1, y_2, z_1, z_2,
y_1', y_2', z_1', z_2') } \\ \nonumber && := - \int d \tilde P_2 \;d \tilde P_3 \:\:
\chi^{P_2}_{\gamma'\beta'}(z_2, y_2)\chi^{P_3}_{\alpha'\gamma}(y_1,
z_1) \:\: \bar\chi^{P_2}_{\beta_1'\gamma_1'}(z_2',
y_2')\bar\chi^{P_3}_{\gamma_1\alpha_1'}(y_1', z_1')\;. 
\end{eqnarray}

Inserting $\tilde G^{(4)}$ and $\tilde G^{(8)}$ into the defining
equation for the irreducible interaction kernel $K^{(6)}$,
see eq. (\ref{eq:K6}), we find the following result for the bound
state contributions to the six--point Green's function: 
\begin{eqnarray}\label{eq:tildeG6K6}
\tilde G^{(6)}_{\alpha\alpha',\beta\beta',\gamma\gamma'}(x_1, x_2,
y_1, y_2, z_1, z_2) &:=& \int d \tilde P_1\; d \tilde P_2 \;d \tilde P_3  \int
d^4x_1' \;d^4x_2' \;d^4y_1' \;d^4y_2' \;d^4z_1' \;d^4z_2'\\ 
&\times& \chi^{P_2}_{\gamma'\beta'}(z_2, y_2)\chi^{P_3}_{\alpha'
\gamma}(y_1, z_1) \:\:\bar \chi^{P_2}_{\beta_1'\gamma_1'}(z_2', y_2')\bar
\chi^{P_3}_{\gamma_1\alpha_1'}(y_1', z_1')\nonumber\\ 
&\times&K^{(6)}_{\alpha_1\alpha_1', \beta_1\beta_1', \gamma_1
\gamma_1'}(x_1', x_2', y_1', y_2' z_1',
z_2')\:\:\chi^{P_1}_{\alpha_1\beta_1}(x_1', x_2')\bar
\chi^{P_1}_{\beta\alpha}(x_1, x_2)\;.\nonumber 
\end{eqnarray}
This expression describes the decay of one initial meson
into two non-interacting mesons; we will now use it to connect the
$S$--matrix element with the irreducible kernel $K^{(6)}$.
 

\subsection{The \boldmath$S$\unboldmath--Matrix Element}
The $S$--matrix operator transforms free states at time $t = - \infty$
into free states at time $t = + \infty$. In terms of a given
interaction Lagrangian ${\cal L}_I$,  the $S$--matrix element
$S_{P_1\rightarrow P_2 P_3}$ for an initial meson with momentum $P_1$
decaying into two outgoing mesons with momenta $P_2$ and $P_3$ can be
written as 
\begin{eqnarray}\label{eq:S-matrix}
S_{P_1\rightarrow P_2 P_3}\; = \; \left\langle P_2 P_3\left|\:\:
\sum^{\infty}_{k=0}\frac{i^k}{k!}\int d^4y_1\cdots d^4y_k \TOP {\cal
L}_I(y_1)\cdots{\cal L}_I(y_k)\right| P_1\right\rangle\;. 
\end{eqnarray}
Here, all fields within ${\cal L}_I$ are taken as free fields and in
normal order. The contribution of this matrix element to the full
six--point Green's function $G^{(6)}$ defined in eq. (\ref{eq:G6}) is given by the expression
\begin{eqnarray}\label{eq:tildeG6}
\tilde G^{(6)}_{\alpha\alpha',\beta\beta',\gamma\gamma'}(x_1, x_2,
y_1, y_2, z_1, z_2) &=& \int d \tilde P_1 d \tilde P_2 d \tilde P_3 \;
\left\langle\: 0\:\left| \TOP \Psi_{\alpha'}(y_1)\bar
\Psi_{\beta'}(y_2)\Psi_{\gamma'}(z_2)\bar\Psi_{\gamma}(z_1)\:\right|\: P_2
P_3 \:\right\rangle\nonumber \\\nonumber 
&\times& \left\langle P_2 P_3\left|
\sum^{\infty}_{k=0}\frac{i^k}{k!}\int d^4y_1\cdots d^4y_k \TOP {\cal
L}_I(y_1)\cdots{\cal L}_I(y_k)\right| P_1\right\rangle\\
&\times& \left\langle \:P_1\:\left|\TOP\Psi_{\beta}(x_2)\bar
\Psi_{\alpha}(x_1)\:\right|\: 0\: \right\rangle\; 
\end{eqnarray}
where now all fields are free fields taken at ${x_1}^0, {x_2}^0
\rightarrow - \infty$ and ${y_1}^0, {y_2}^0, {z_1}^0, {z_2}^0 \rightarrow
+\infty$. As the state $|P_2 P_3 \rangle$ is a product of two
non--interacting meson states, \ie $|P_2 P_3\rangle = |P_2\rangle |P_3
\rangle$, the field operators with the quantum numbers $\alpha',
\gamma$ act (by convention) only on the state $| P_3\rangle$ and the
field operators labeled with $\beta', \gamma'$ act only on the state
$|P_2\rangle$. Thus we can write 
\begin{eqnarray}
\big\langle \:0\:\big|\: \TOP \Psi_{\alpha'}(y_1)\bar
\Psi_{\beta'}(y_2)\Psi_{\gamma'}(z_2)\bar\Psi_{\gamma}(z_1)\:\big|
\:P_2 P_3 \:\big\rangle = - \chi^{P_3}_{\alpha'\gamma}(y_1,
z_1)\:\chi^{P_2}_{\gamma'\beta'}(z_2, y_2) 
\end{eqnarray}
and therefore eq. (\ref{eq:tildeG6}) becomes
\begin{eqnarray}
\tilde G^{(6)}_{\alpha\alpha',\beta\beta',\gamma\gamma'}(x_1, x_2,
y_1, y_2, z_1, z_2)  = \int d \tilde P_1 d \tilde P_2 d \tilde P_3
\;\:\: \chi^{P_2}_{\gamma'\beta'}(z_2, y_2)
\chi^{P_3}_{\alpha'\gamma}(y_1, z_1) \:\:S_{P_1\rightarrow P_2
P_3}\:\:\bar \chi^{P_1}_{\beta\alpha}(x_2, x_1)\;. 
\end{eqnarray}

By comparing this expression with eq. (\ref{eq:tildeG6K6}), the
connection between the $S$--matrix element $S_{P_1\rightarrow P_2
P_3}$, the irreducible kernel $K^{(6)}$ and the Bethe-Salpeter
amplitudes of the three mesons can be written according to
\begin{eqnarray}\label{eq:SmatrixK6}
S_{P_1\rightarrow P_2 P_3} &=& - \int d^4x_1' d^4x_2' d^4 y_1' d^4y_2'
d^4z_1' d^4z_2' \; \:\:\\ \nonumber &\times& \tr\Big[\bar
\chi^{P_2}(z_2', y_2')\:\bar\chi^{P_3}(y_1', z_1') 
\:\: K^{(6)}(x_1', x_2', y_1', y_2', z_1', z_2')
\:\:\chi^{P_1}(x_1', x_2')\Big]\;; 
\end{eqnarray}
here, we have abbreviated the full contraction of all multi--indices
by the trace symbol $\tr$. In the following, we will specify the
particular interactions summarized in the six--point kernel $K^{(6)}$.


\subsection{Approximation of the Interaction Kernel}
The uncontracted term of the 't~Hooft Lagrangian $\Delta {\cal
L}^{\mbox{\scriptsize eff}}$ (see appendix \ref{app:tHooft}) yields an explicit
expression for an effective six--quark interaction Lagrangian ${\cal
L}_I = {\cal L}^{(3)}$ given in eq. (\ref{L3}). Up to the first
order in the instanton coupling $g^{(3)}_{\mbox{\scriptsize eff}}$, the
$S$--matrix element $S_{P_1\rightarrow P_2 P_3}$ in
eq. (\ref{eq:S-matrix}) consists of two terms: 
\begin{eqnarray}
\label{SMatrixElement}
S_{P_1 \rightarrow P_2 P_3} = \big\langle \: P_2 P_3\: \big|\: P_1\: \big\rangle +
\big\langle\:  P_2 P_3\: \big|\: \:  i \int d^4 y \: \: {\cal L}^{(3)}
(y)\: \: \big|\: P_1\: \big\rangle + {\cal O}\Big((g^{(3)}_{\mbox{\scriptsize eff}})^2\Big)\quad. 
\end{eqnarray}
The first term is of order ${\cal O}(1)$ in the coupling of the
six--quark interaction, \ie the 't~Hooft three--body
interaction via ${\cal L}^{(3)}$ is absent and
the decay passes exclusively over the propagation of non-interacting
quarks. The second term is of order ${\cal O}(g^{(3)}_{\mbox{\scriptsize eff}})$,
\ie the quarks of the three mesons once interact via
${\cal L}(3)$.  

To find the corresponding interaction kernel, we consider the six--point
Green's function, see eq. (\ref{eq:G6}), up to this
order; it thus can be decomposed into two terms accordings to
$G^{(6)} = G^{(6)}_0 + G^{(6)}_1 + \ldots$. 
In the following, the (anti--)quark fields $\Psi (\bar \Psi)$ are free
fields in the sense that they are out of the interaction region. By
using Wick's theorem and after inserting several
$\delta$--distributions, we get for the lowest--order term the expression 
\begin{eqnarray}
\label{eq:G60delta}
\lefteqn{G^{(6)}_{0\:\alpha\alpha',\beta\beta',\gamma\gamma'}(x_1, x_2,
y_1, y_2, z_1, z_2) := \big\langle\: 0 \:\big| \TOP \Psi_{\alpha'}(y_1)\bar
\Psi_{\beta'}(y_2)\Psi_{\gamma'}(z_2)\bar
\Psi_{\gamma}(z_1)\Psi_{\beta}(x_2)\bar\Psi_{\alpha}(x_1)\:\big|\:0\:\big\rangle\nonumber
}\\
=& -& \int d^4x_1'd^4x_2'd^4y_1'd^4y_2'd^4z_1'd^4z_2' \:\:
\delta_{\alpha_1\alpha'}\delta_{\beta_1\beta'}\delta_{\gamma_1\gamma'}\nonumber
\\ 
&\times& \delta(x_1'- y_1')\:\:\delta(x_2' -y_2')\:\:\delta(z_1' -
z_2')\:\:\delta(y_1' - y_1)\:\:\delta(y_2' - y_2)\:\:\delta(z_2' - z_2)\nonumber\\ 
&\times&\big\langle\: 0\:\big| \TOP \Psi_{\alpha_1}(x_1')\bar \Psi_{\alpha}(x_1)\:\big|\:
0\:\big\rangle \:\big\langle\:
0\:\big|\TOP\Psi_{\beta}(x_2)\bar\Psi_{\beta_1}(x_2')\:\big|\:0\:\big\rangle\:\big\langle
0\:\big|\TOP\Psi_{\gamma_1}(z_1')\bar\Psi_{\gamma}(z_1)\:\big|\:0\:\big\rangle\nonumber\\ 
&+& \mbox{crossed term}\;\:\:.\nonumber
\end{eqnarray}
The crossed term follows from the first term by exchanging 
($z_2, \gamma'$)$\longleftrightarrow$($y_1, \alpha'$) and ($z_1,
\gamma$)$\longleftrightarrow$($y_2, \beta '$); see also
figs.~\ref{fig:FULLKSIX} and \ref{fig:SDloopinst}.  

Now we approximate the full Feynman propagators by the free
propagators with effective constituent quark masses, consistent with
the approximation done in the Bethe-Salpeter equation (see section
\ref{sect:BSapproach}). In coordinate space, they satisfy the relations
\begin{eqnarray}
(i \not \!\partial_x - m)_{\alpha\alpha'}S^F_{\alpha'\beta}(x, x') &=& +
i \delta(x-x')\delta_{\alpha\beta}\\ \mbox{and}\quad
S^F_{\alpha\alpha'}(x, x') (i \not \!\partial_{x'} + m)_{\alpha'\beta}
&=& - i \delta(x-x')\delta_{\alpha\beta}\quad .
\end{eqnarray} 

Inserting these relations into eq. (\ref{eq:G60delta}) and using
$S^F_{\alpha\alpha'}(x, x'):=\langle 0\:|\TOP\Psi_{\alpha}(x)\bar \Psi_{\alpha'}(x')\:|\:0\rangle$, we
get for the quark loop part $G^{(6)}_0$ of the six--point Green's
function the following result:
\begin{eqnarray}
G^{(6)}_{0\:\alpha\alpha',\beta\beta',\gamma\gamma'}(x_1, x_2,
y_1, y_2, z_1, z_2) &=& - i \int d^4x_1' d^4x_2' d^4y_1' d^4y_2' d^4z_1' d^4z_2'\nonumber \\
&\times& \Big[ S_1^F(y_1,y_1')\big(-i \not \!\partial_{y_1'} - m_1\big)
S_1^F(x_1',x_1)\Big]_{\alpha '\alpha}\:\: \delta (x_1' - y_1') \nonumber\\ 
&\times& \Big[ S_2^F(x_2,x_2')\big(+i \not \!\partial_{y_2'} - m_2\big)
S_2^F(y_2',y_2)\Big]_{\beta   \beta'}\:\: \delta (x_2' - y_2') \nonumber\\ 
&\times& \Big[ S^F_3(z_2,z_2')\:\big(-i \not \!\partial_{z_2'} - m_3\big)
\:S^F_3(z_1',z_1)\Big]_{\gamma '\gamma} \:\: \delta (z_2' - z_1') \nonumber \\
&+&\mbox{crossed term}\;.
\end{eqnarray}
Here, the constituent quark masses of the incoming quark and anti--quark
are denoted by $m_1$ and $m_2$, respectively; the constituent quark
mass of the third quark is labeled by $m_3$. 

Now we want to determine the irreducible six--point kernel $K^{(6)}$
introduced in eq. (\ref{eq:K6}). In this defining equation, we now
approximate the four--point Green's function by its free part and
adopt $G^{(4)}_{\alpha_1\alpha,\beta_1\beta}(x_1',x_2',x_1,x_2)\approx
S^F_{\alpha_1\alpha}(x_1',x_1)S^F_{\beta\beta_1}(x_2,x_2')$. Accordingly,
we insert this four--point Green's function in the decomposition
$G^{(8)} \approx G^{(4)}\cdot G^{(4)}$ of the eight--point Green's
function since we neglect all effects originating in final state
interactions of the outgoing mesons. With this assumptions, we can read
off the lowest--order part of the six--point kernel defined in
eq. (\ref{eq:K6}) as follows:

\equ{
K^{(6)}_{0\:\alpha_1\alpha_1', \beta_1\beta_1', \gamma_1\gamma_1'}(x_1',
x_2', y_1', y_2', z_1', z_2')  = &-i& \: \delta (x_1' - y_1')\: \delta
(x_2' - y_2')\: \delta (z_2' - z_1')\\ 
&\times&\big(-i \not \!\partial_{y_1'} - m_1\big)_{\alpha_1'\alpha_1}\:\big(+i \not
\!\partial_{y_2'} - m_2\big)_{ \beta_1\beta_1'}\:\big(-i \not \!\partial_{z_2'} -
m_3\big) _{\gamma_1'\gamma_1}\nonumber \\
&+&\mbox{crossed term} \nonumber\quad .
}

The next term  $G^{(6)}_1$ in the six--point Green's function is of
order ${\cal O}(g^{(3)}_{\mbox{\scriptsize eff}})$  in the
instanton--induced interaction; starting with its definition 

\equ{
\lefteqn{G^{(6)}_{1\:\alpha\alpha',\beta\beta',\gamma\gamma'}(x_1, x_2,
y_1, y_2, z_1, z_2) }\\ &:=& \nonumber i \int d^4 y \big\langle\:
0\:\big|\TOP\Psi_{\alpha'}(y_1)\bar
\Psi_{\beta'}(y_2)\Psi_{\gamma'}(z_2)\bar
\Psi_{\gamma}(z_1)\Psi_{\beta}(x_2)\bar\Psi_{\alpha}(x_1) \: {\cal L}^{(3)}
(y)\:\big|\:0 \:\big\rangle \:,
}

it can be derived in analogy to the
pure quark loop part. Using the anti--symmetry of the operator ${\cal
O}^{FSC}$ defined in eq. (\ref{O_FSC}) in appendix \ref{tHooft3Body}, the result for
the interaction kernel reads 

\equ{
\lefteqn{K^{(6)}_{1\:\alpha_1\alpha_1', \beta_1\beta_1', \gamma_1\gamma_1'}(x_1',
x_2', y_1', y_2', z_1', z_2')  = -i \: \; 36 \;
g^{(3)}_{\mbox{\scriptsize eff}}\; \cdot \;{\cal
O}^{FSC}_{\beta_1\alpha_1'\gamma_1'\alpha_1\gamma_1\beta_1'}}\\  
&\times& \; \int d^4y \:\:\delta(x_1' - y)\:\delta(x_2' -y)\:\delta(y_1' -
y)\:\delta(y_2' - y)\:\delta(z_1' - y)\:\delta(z_2' - y) \nonumber\quad .
}

where we have inserted the explicit expression for the
instanton--induced interaction Lagrangian ${\cal L}^{(3)}$ (see appendix
\ref{app:tHooft}); the numerical factor 
arises from $36$ identical terms found by performing Wick's theorem to
$G^{(6)}_1$.  By comparison of $G^{(6)} = G^{(6)}_0 + G^{(6)}_1$ up to
order ${\cal O}(g^{(3)}_{\mbox{\scriptsize eff}})$ with the
defining equation (\ref{eq:K6}) for the irreducible interaction
kernel, we thus find $K^{(6)}=K^{(6)}_0+K^{(6)}_1$ in the same
approximation.


\subsection{The Transition Matrix Element}

The $S$--matrix element as given in
eq. (\ref{eq:SmatrixK6}) up to first order in the three--body coupling
constant $g^{(3)}_{\mbox{\scriptsize eff}}$  follows from insertion of
the irreducible six--point kernel in the same order. To obtain a more
compact notation, we relabel some integration variables, use again the
trace symbol $\tr$ for the contraction of all multi--indices and
finally find

\begin{eqnarray}
S_{P_1\rightarrow P_2 P_3} 
&=& i \int d^4x \:d^4 y \:d^4z\: \tr\Big[\bar \chi^{P_2}(z,y)\big(i \not \!\partial_{z} + m_3\big)\bar
\chi^{P_3}(x,z)\big(i \not \!\partial_{x} +
m_1\big)\chi^{P_1}(x,y)\big(i \not \!\partial_{y} -
m_2\big)\Big]\nonumber\\  
&+& i \int d^4x \:d^4 y \:d^4z\: \tr\Big[\bar \chi^{P_3}(z,x)\big(i \not \!\partial_{z} + m_3\big)\bar
\chi^{P_2}(y,z)\big(i \not \!\partial_{y} +
m_1\big)\chi^{P_1}(y,x)\big(i \not \!\partial_{x} -
m_2\big)\Big]\nonumber\\  
&+& i \; 36\; g^{(3)}_{\mbox{\scriptsize eff}}\; \int d^4 y \:\tr
\Big[\:{\cal O}^{FSC}\:\big(\bar\chi^{P_3}(y, y)\otimes\bar\chi^{P_2}(y,
y)\otimes\chi^{P_1}(y, y)\big)\:\Big]\;. 
\end{eqnarray}

We now perform a Fourier transformation into momentum space and use the
well--known general relation between the transition operator $T$ and the $S$--matrix
operator given by $S_{f i} = \delta_{f i} + i (2\pi)^4 \delta(P_f - P_i) \langle f|T| i \rangle $
with $|i\rangle$ and $|f\rangle$ initial and final state,
respectively. Finally, we find the following transition amplitude
 up to first order in the coupling of
't~Hooft's instanton--induced three--body interaction: 

\begin{eqnarray}
\label{TransitionMatrixElement}
T_{P_1\rightarrow P_2 P_3} &=&  \big\langle\: P_2 P_3\:\big|\: T \:|\:P_1\:\big\rangle=
T_{P_1\rightarrow P_2 P_3}^{\mbox{\scriptsize loop}} + 
T_{P_1\rightarrow P_2 P_3}^{\mbox{\scriptsize 't~Hooft}} + {\cal
O}\Big((g^{(3)}_{\mbox{\scriptsize eff}})^2\Big)\\[2ex] \nonumber
\mbox{with}\:\: T_{P_1\rightarrow P_2 P_3}^{\mbox{\scriptsize loop}}
&:=& \int \frac{d^4p}{(2\pi)^4}\: \tr \Big[ 
\bar\Gamma^{P_2}(p-\frac{P_3}{2}) S^F_3(\frac{P_2-P_3}{2}+p)
\bar\Gamma^{P_3}(p+\frac{P_2}{2}) S^F_1(\frac{P_1}{2} + 
p)\Gamma^{P_1}(p) S^F_2(-\frac{P_1}{2} + p)\Big]\\  \nonumber
&+& \int \frac{d^4p}{(2\pi)^4}\: \tr \Big[
\bar\Gamma^{P_3}(p-\frac{P_2}{2}) S^F_3(\frac{P_3-P_2}{2}+p)
\bar\Gamma^{P_2}(p+\frac{P_3}{2}) S^F_1(\frac{P_1}{2} + 
p)\Gamma^{P_1}(p) S^F_2(-\frac{P_1}{2} + p)\Big]\\[1ex]
\mbox{and}\:\: T_{P_1\rightarrow P_2 P_3}^{\mbox{\scriptsize 't~Hooft}} 
&:=& \:\:36 \: g^{(3)}_{\mbox{\scriptsize eff}} \:\:\tr \:\Big[\:{\cal O}^{FSC}\:
\big(\:\bigotimes_{i=1}^3 \int \frac{d^4p_i}{(2\pi)^4}\: {\cal
X}_i^{P_i}(p_i)\:\big)\:\Big] \quad .\nonumber
\end{eqnarray}

The trace runs over colour, flavour and spin indices;
by definition, momentum conservation is fulfilled, \ie $P_1 = P_2 +
P_3$ holds. In the quark loop part, we have expressed the
Bethe-Salpeter amplitudes by the corresponding vertex functions
according to their definition given in section \ref{sect:BSapproach}. 
Note that we have
introduced ${\cal X}_1=\chi$ and ${\cal X}_2={\cal X}_3=\bar\chi$
in order to distinguish properly the Bethe--Salpeter amplitude and its adjoint. 


\subsection{Calculation of the Decay Width}
\label{CalculationOfTheDecayaWidth}

After we have derived the transition matrix element $T_{P_1\rightarrow
P_2 P_3}$ for the mesonic strong decay $\123$, we will finally
give the standard formula for the calculation of the partial decay
width:
\begin{eqnarray}
\label{DecayWidthFormula}
\Gamma_{\123} &=& \:\frac{k}{8\pi
  M^2_1}\:\:\sum_{m_{1}} \frac{1}{2J_1+1}\sum_{m_{2},m_{3}}\Big|\big\langle \:P_2 m_{2}\: P_3
  m_{3}\:|\: T\:|\:P_1 m_{1}\:\big\rangle\Big|^2 \quad ;
\end{eqnarray}

here, $k^2 = (P^0_2)^2 - M^2_2 = (P^0_3)^2 - M^2_3$ in the rest frame
of the incoming meson (\ie $P_1={M_1\choose \vec 0}$) is a kinematical
factor, $m_i:= m_{J_i}$ ($i=1,2,3$) are the magnetic quantum numbers
and $P_1 = P_2 + P_3$ holds due to four--momentum 
conservation. In appendix \ref{app:FlavourFactors}, we comment on the
flavour matrix element and on the particular factors corresponding to
the charge multiplicities.

\begin{figure}
\begin{center}
\input {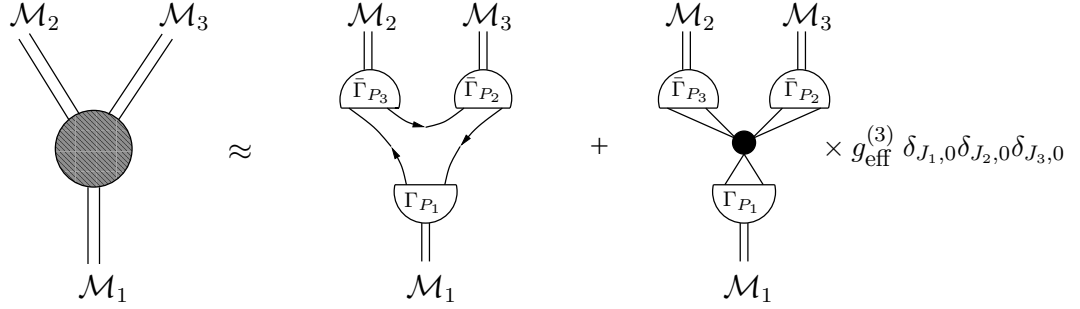}
\caption{{\small Diagrammatic illustration of the decay of one meson
with four momentum $P_1$ into two other mesons with four momenta $P_2$
and $P_3$ via the full transition matrix element including quark loops
and the instanton--induced six--quark vertex; note that the latter
only contributes if all mesons in the decay have zero angular
momentum. The quark loop diagram 
with the two outgoing mesons exchanged as well as the cyclic
permutated diagrams of the instanton--induced transitions are
suppressed in this figure.}}\label{fig:SDloopinst} 
\end{center}
\end{figure}

In fig.~\ref{fig:SDloopinst}, we summarize the results of this section
in a diagrammatical representation. As we have shown in the framework
of the Mandelstam formalism, the pure quark loop term contributes to
all strong two--body decays. In contrast to this general mechanism,
the instanton--induced decay via the three--body vertex of the 't
Hooft Lagrangian only takes place if all mesons in the decay are either
of scalar or pseudoscalar type, \ie if $J_i=0$ ($i=1,2,3$) holds. In this
case, we expect the interference between the quark loop terms and the
instanton--induced contributions to modify the results on the strong
decay widths in our calculation in a characteristical fashion.

Let us make a final remark before we study in detail the numerical
results concerning the strong meson decays. It is obvious from
fig.~\ref{fig:SDloopinst} that the quark loop part satisfies the
well--known Zweig rule (or OZI rule). For the instanton--induced decay
mechanism, the situation is different if and only if a flavour
singlet participates in the decay under consideration. In order to see
this, we write down the explicit flavour dependence of this interaction
(see also \cite{RitterPaper,RitterDiplom,RickenPhD}) as 
\label{OZIViolation}
\begin{eqnarray}\label{eq:sixquarkfldep}
\tr \Big[\:{\cal P}^{F}_1(\Lambda^{{\cal M}_1}\otimes\Lambda^{{\cal
M}_2}\otimes\Lambda^{{\cal M}_3})\:\Big]\;=\;\frac 1 6
\epsilon^{ijk}\epsilon^{i'j'k'}\Lambda^{{\cal
M}_1}_{ii'}\Lambda^{{\cal M}_2}_{jj'}\Lambda^{{\cal M}_3}_{kk'}\;. 
\end{eqnarray} 
Here, $\Lambda^{{\cal M}_i}$ ($i=1, 2, 3$) is the flavour
part of Bethe--Salpeter amplitude for the meson ${\cal M}_i$ and the
whole expression originates in the term $T_{P_1\rightarrow P_2
P_3}^{\mbox{\scriptsize 't~Hooft}}$ of
eq. (\ref{TransitionMatrixElement}). Note that ${\cal P}^{F}_1$
projects  onto singlet states in flavour space; this
projector is part of the general operator ${\cal O}^{FSC}$ defined in
eq. (\ref{O_FSC}) in appendix \ref{tHooft3Body}. The
expression in eq. (\ref{eq:sixquarkfldep}) shows that the
instanton--induced six--quark interaction is completely antisymmetric
in flavour space. With the help of the Cayley-Hamilton theorem, the
flavour dependence can be rewritten as 
\begin{eqnarray}\label{eq:sixquarkfldep2}
\tr\Big[\:{\cal P}^{F}_1(\Lambda^{{\cal M}_1}\otimes\Lambda^{{\cal
M}_2}\otimes\Lambda^{{\cal M}_3})\:\Big] &=& \frac 1 6 \tr\Big[\:\Lambda^{{\cal
M}_1}\Lambda^{{\cal M}_2}\Lambda^{{\cal M}_3}+\Lambda^{{\cal
M}_1}\Lambda^{{\cal M}_3}\Lambda^{{\cal M}_2}\:\Big] \\ &-& \frac 1
6\Big(\:\tr\Big[\:\Lambda^{{\cal M}_1}\Big]\:\tr\Big[\:\hat
\Lambda^{{\cal M}_2} \hat 
\Lambda^{{\cal M}_3}\:\Big] + \mbox{ cycl. perm.}\:\Big)\nonumber 
\end{eqnarray} 
where $\hat \Lambda^{{\cal M}_i}$ denotes the traceless part of
$\Lambda^{{\cal M}_i}$ (see \cite{RitterPaper,RitterDiplom}). Here,
the first term has the flavour dependence 
of usual quark loop diagrams --- see fig.~(\ref{fig:SDloopinst}) ---  such
that this interaction gives an additional contribution to the
conventional Zweig rule allowed transitions for decays involving only
$J_i=0$ mesons ($i=1, 2, 3)$. The flavour dependence of the
second term leads to a minimal violation of the OZI rule: only if
$\tr [\Lambda^{{\cal M}_i}]$ does not vanish, \ie if and only if a flavour
singlet participates, there is an additional contribution to the
conventional decay mechanism beyond the Zweig rule allowed
processes.


\section{Results and Discussion}
\label{Results}

As we have already mentioned in section \ref{IntParSpec}, we have
calculated the masses of light mesons in two different parameter sets for the 
confinement force (see \cite{KollRicken,RickenKoll}). Both models
yielded an excellent description of 
the experimental ground state Regge trajectories while the masses of the
pseudoscalar  ground state nonet were very well described by using 't
Hooft's force as an additional residual interaction. However, we found
the radial excitation spectrum as well as the complete scalar meson
spectrum considerably different in both models. Especially the latter
sector needs further investigations beyond purely
spectroscopical considerations in order to understand the nature of
the scalar mesons. The study of hadronic decays presented in this
contribution may help for a more serious interpretation of these
mesons. However, the calculation of strong decay widths for the complete
resonance spectrum is of course first of all important in its own
right. 

In this publication, we study strong two--body decays under the assumption
that any final state interaction can be neglected. In this
approximation, the transition matrix element 
for the decay of one initial meson ${\cal M}_1$ with four momentum $P_1$ 
into two outgoing mesons ${\cal M}_2$ and ${\cal M}_3$ with four momenta $P_2, P_3$ (for the bound
states masses, $M_i^2=P_i^2$ holds) is
evaluated up to first order in the effective coupling of
the instanton induced three--body interaction ${\cal L}^{(3)}$, see
eqs. (\ref{SMatrixElement}) and (\ref{TransitionMatrixElement}). 

In the lowest order term, the instanton--induced interaction is
absent and the decay passes over the propagation of non--interacting
quarks forming a simple loop. These transitions are
Zweig rule allowed and they always contribute, independent of the
quantum numbers of the three mesons that are involved in the decay
under consideration. The next term in the transition matrix element
of eq. (\ref{TransitionMatrixElement}) is of order ${\cal
O}(g^{(3)}_{\mbox{\scriptsize eff}})$, \ie the quarks of the three mesons
once interact via the instanton--induced interaction described by ${\cal
L}^{(3)}$ (see also appendix \ref{tHooft3Body}). These transitions
only contribute if all three mesons have 
vanishing total angular momentum $J_i=0$ $(i=1,2,3)$ such that the decays of other
than scalar and pseudoscalar mesons are not influenced by this
interaction. The transition amplitude generated by ${\cal L}(3)$
contains a Zweig rule violating part if a flavour singlet meson is
involved in the decay under consideration.\footnote{This hierarchy of
decay mechanisms in our approach suggests a clear 
indication whether or not the instanton--induced effects contribute to
a specific decay. In all tables summarizing our numerical
results, we thus indicate by the symbol ``$\bullet$'' in the column
``$III$'' if the three--body 't Hooft interaction
plays a role, \ie if we expect an interference of the amplitudes originating both
in quark loop diagrams and instanton--induced decay
mechanisms. Furthermore, we point out by an additional symbol
``$\bullet$'' in the column ``ZRV'' whether
Zweig rule violating processes contribute to the decay width or not;
as we have shown above, effects beyond the OZI rule allowed diagrams
are expected if isoscalar mesons participate the instanton--induced decay process.}

At this point, a short remark on our selection of the decays in our
following discussion is in order.\footnote{Note that we denote the
final state ``$K\bar K+c.c.$'' by ``$KK$''; analogously,
we abbreviate the ``$K\bar K^*+c.c.$'' final state with ``$KK^*$'' for simplicity.}
In general, we only quote our results on 
particular decay widths if it can either be compared with an
experimental value or with data obtained in the framework of the
$^3P_0$ model. For the latter reference, we usually quote the results of
T.~Barnes \etal published in
refs. \cite{BarnesStrongDecays1,BarnesStrongDecays2}; we refer to
R.~Bonnaz and B.~Silvestre-Brac with respect to the decays of
non--$n\bar n$ mesons (see \cite{BonnazSilvestreBrac}).\footnote{To
be precise, we refer to their parameter set labeled ``NRAL'' with a
momentum--dependent $^3P_0$--vertex including instanton--induced as
well as tensor--force effects. For outgoing mesons with broad widths
($> 50$~MeV), the authors introduced a modified mechanism for the
description of this situation; note that this approach has severe
problems due to the inherent violation of Galilei invariance of the
$^3P_0$ model.}

Let us give also a brief comment on two recently published works: In
\cite{Jarecke:2002xd},  strong decays of light vector mesons are
studied within a covariant approach combining Bethe-Salpeter and
Dyson-Schwinger equations. For the related coupling constants the
authors find good agreement with the experimental results; note that
only quark loops contribute to their calculated widths. A further
interesting review (see \cite{Barnes:2002mu}) describes $^3P_0$ model
results for strange quarkonia. As most of the strong decay widths in
this paper are parametrized in terms of a singlett--triplett mixing
angle, which is not uniquely fixed {\sl a priori} by experiment, we
will not discuss their results in more detail.


\subsection{Fixing the Three--Body Coupling Constant}\label{Fixing_g3}

\begin{figure}
\begin{center}
\input {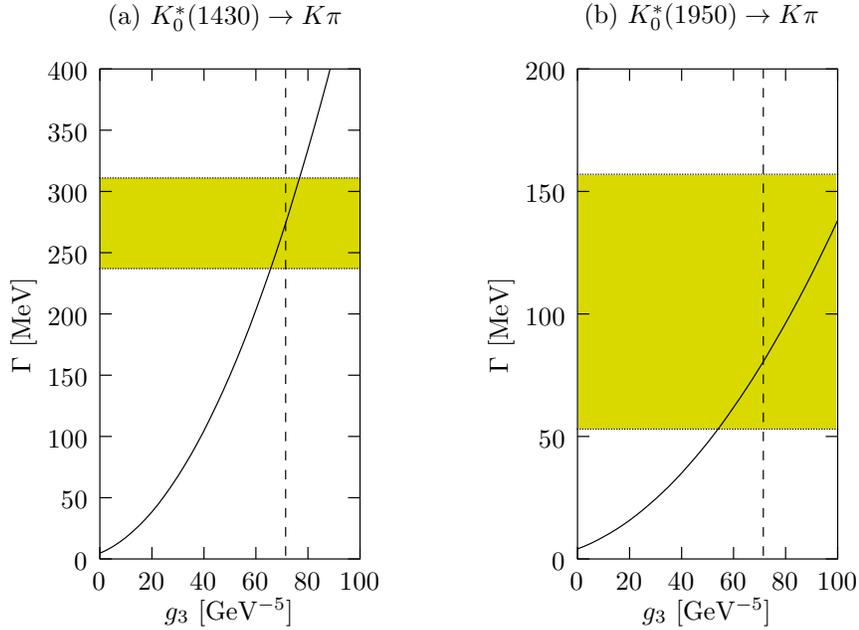}
\caption{{\small The widths of the decays (a) $K_0^*(1430)\to K\pi$ and
(b) $K_0^*(1950)\to K\pi$ and their dependence on the coupling constant
$g^{(3)}_{\mbox{\scriptsize eff}}$ of 't~Hooft's three--body force due
to instanton effects; see text for more details. The shaded bands
denote the experimental error bars according to the PDG (see
\cite{PDG2000}); the vertical dashed lines mark our choice for the
't~Hooft coupling $g^{(3)}_{\mbox{\scriptsize eff}}$.}}
\label{fig:Fixing_g3} 
\end{center}
\end{figure}

Before we discuss the numerical results for the various decay widths,
we shall briefly comment on our choice for the value of the coupling
constant $g^{(3)}_{\mbox{\scriptsize eff}}$ which governs the strength
of the instanton--induced six--quark vertex. As we show in
fig.~\ref{fig:Fixing_g3}(a), we fix this free parameter at the
width for the decay $K_0^*(1430)\to K\pi$, namely
$\Gamma_{\mbox{\small exp}}=274\pm 37$~MeV. Compared to other decays in
which all participating mesons have total angular momentum 
$J_i=0$, this width is quite well--known experimentally which suggests
to fix the free parameter by this number. However, the uncertainty is
still large such that we cannot be sure to have optimally fixed
$g^{(3)}_{\mbox{\scriptsize eff}}$. 

Moreover, the parameter set used
in this work has the special feature that --- at a first glance ---
underestimates the masses of the scalar bound states such as the
$K_0^*$ mesons; note that nevertheless 
our results agree excellently with the $K$--matrix analyses done by
V.~V.~Anisovich and co--workers (see \cite{Anisovich} as well as our
discussion in \cite{RickenKoll}). Thus, the particular phase
space factor in the decay $K_0^*(1430)\to K\pi$ is not well described
in our model as the scalar kaon mass is roughly 200~MeV smaller than
compared to the standard value of the PDG (see \cite{PDG2000}). We
have however decided to fix the 't~Hooft coupling
$g^{(3)}_{\mbox{\scriptsize eff}}$ at the experimental value for the
decay $K_0^*(1430)\to K\pi$. Our approach is justified by the
observation that our choice simultaneously allows for a correct
description of the decay $K_0^*(1950)\to K\pi$, see
fig.~\ref{fig:Fixing_g3}(b). In particular, it is impressing to see
how the instanton--induced contribution to the decay widths in these
examples raises dramatically the pure quark loop contribution of only
a few MeV up to the experimental values of several hundred MeV.


\subsection{Decays of \boldmath$\rho _J$\unboldmath\/ Mesons}\label{rhoJ_SUBSECTION}

In tab.~\ref{rho}, we have listed our result concerning the strong
two--body decays of isovector $\rho _J$ mesons; in particular, we study the
$\rho(770)$, $\rho(1450)$, $\rho(1700)$ and $\rho_3(1690)$. We shall
note that the radial excitation pattern in this sector is slightly
different to what is usually assumed for the $\rho$ mesons: while the
masses of the $\rho(770)$ and $\rho(1450)$ are quite well described in
our approach, we underestimate the $\rho(1700)$ mass by roughly
$\approx 200$~MeV (see \cite{KollRicken}). Note that our calculation
implies that the $\rho(770)$ and $\rho(1700)$ are dominantly $S$--wave
mesons while the $\rho(1450)$ and $\rho_3(1690)$ are dominantly
$D$--wave mesons (see appendix of ref. \cite{KollPhD} for comparison).

Comparing our result for the $\rho(770)\to\pi\pi$ decay with
experiment and the $^3P_0$ model, we see that both theoretical
approaches underestimate the partial decay width $\Gamma_{\mbox{\small
exp}}\approx 150$~MeV such that the strong two--body decay mode alone
cannot saturate the total $\rho(770)$ width in both approaches. This
is a first hint that the neglect of final--state interaction in our
ansatz might spoil our results in certain sectors although we use a
completely relativistic formula for the calculation of the quark loop
diagram. 

The various decay modes of the $\rho(1450)$ are in general very small
in our approach compared to the $^3P_0$ model. However, a look at the
experimental limits in this sector leads to the conclusion that our
results for the decays $\rho(1450)\to\omega\pi/\rho\eta/KK$ are at
least consistent with these rough limits while the results of
ref. \cite{BarnesStrongDecays2} strongly overestimate the related
partial widths. For the $\rho(1700)$, an analogous statement holds:
the widths calculated in the framework of our model are very small
compared to the $^3P_0$ --- however, they do not contradict the
experimental limit \eg for the decay $\rho(1700)\to\rho\eta$ with
$\Gamma_{\mbox{\small exp}}< 9.6$~MeV. For both radial $\rho$
excitations, the total experimental width is again significantly larger
than the sum of the partial width calculated in our approach.

For the strong decays of the $\rho_3(1690)$, we find that especially
the $\pi\pi$ partial width is significantly too small. This observation
might be related to the fact that the description of the pion in the
framework of our model is not overall satisfying; as we have already
noted in ref. \cite{KollRicken}, the instantaneous approximation seems
to be not well suited for a deeply bound state such as the $\pi$
meson. For the strong decays, the related wave function deformation
leads to a general underestimation of partial widths if one or two
pions are observed in the final state. As our calculated widths for the
decays $\rho_3(1690)\to\rho\eta/KK$ are plausible compared to
experiment as well as to the $^3P_0$ model result, this problem seems
to be restricted to pionic final states at the moment. In fact, the large
 $\rho_3(1690)\to\rho\rho$ width in our framework  is comparable with the $^3P_0$ model.


\subsection{Decays of \boldmath$\pi _J$\unboldmath\/ Mesons}\label{piJ_SUBSECTION}

The results of the strong $\pi _J$ meson decay are presented in
tab.~\ref{pi}; let us note that the masses of the $\pi(1300)$,
$\pi(1800)$ and $\pi_2(1670)$ are well described in our model
(see \cite{KollRicken}).

Again, we find that the calculated widths are significantly smaller
than the results of the $^3P_0$ model but the experimental data are
quite poor for the $\pi(1300)$ and $\pi(1800)$ such that a detailed
comparison is difficult. We note that we also find small widths for
the decays $\pi(1800)\to\rho\pi/KK^*$ which have not been seen in
experiment; here, the approach of ref. \cite{BarnesStrongDecays2} seems
to overestimate the partial widths.

At this point, a further remark concerning the contributions of
't~Hooft's six--quark interaction is in order. As we have denoted it
in tab.~ \ref{pi} in the columns ``III'' and ``ZRV'', the decays
$\pi(1800)\to f_0(1370)\pi$/$f_0(1500)\pi$/ $a_0(980)\eta$/$KK_0^*$ not only
include the simple quark loop mechanism but also obtain a
significant contribution by the instanton induced interaction. In
general, the interference between both terms can either be
constructive or destructive. For the pion decays discussed in this
section, the interference is destructive and therefore lowers the
resulting partial widths compared to the pure quark loop contribution.

Concerning the $\pi_2(1670)$ decays, we still find generally too small
widths; at least for the pion--free final state $KK^*$, we obtain a
plausible result. Note that the $^3P_0$ model gives a vanishing
partial width for the $f_0(1370)\pi$ decay mode in contradiction to
experiment; also for other final states, the non--relativistic $^3P_0$
ansatz is less accurate than often claimed.


\subsection{Decays of \boldmath$\omega _J$\unboldmath\/ and
\boldmath$\phi _J$\unboldmath\/ Mesons}\label{omegaphiJ_SUBSECTION} 

In tab.~\ref{omegaphi}, we summarize our results for the strong
partial decay widths of the isoscalar $\omega_J$ and $\phi_J$
mesons. Their masses are quite well described in our Bethe--Salpeter
ansatz; note that --- due to the lack of a mixing mechanism for
$J\not=0$ mesons --- the $\omega_J$ are purely $n\bar n$ states while
the $\phi_J$ are purely $s\bar s$ states in good agreement with their
experimental status. We assume the $\omega(1420)$ to be the second
radial $n\bar n$ excitation in this sector as this meson is considered
as an $S$--wave state in ref. \cite{BarnesStrongDecays2}; analogously,
we consider the $\phi(1680)$ to  be the second radially excited $s\bar
s$ state (see also the appendix of ref. \cite{KollPhD}).  

As we have already observed before, the total widths of the $\omega_J$
mesons cannot be saturated by the sum of their two--body strong decay
widths alone; although there in fact exist numerous other
(pre--dominantly electromagnetic) decay modes (see \cite{PDG2000}),
this leads again to some inconsistencies between our relativistic quark
loop approach and the results of the $^3P_0$ model.

Unfortunately, we cannot compute the $\phi(1020)\to KK$ decay due to
kinematical reasons: although the calculated masses for the $\phi$ and
the $K$ mesons are accurate up to only $\approx 3$\%, the relation
$M_\phi\ge 2M_K$ is missed by a few MeV and thus a calculation of this
decay is prohibited. We could of course slightly re--adjust our model
parameters (\eg to lower the $K$ mass a little bit); however, we have
refrained from such a manipulation as the decay rate is obviously
extremely sensible on the precise values of the phase space factor in
$\phi(1020)\to KK$, see also eq. (\ref{DecayWidthFormula}). The other
partial widths of the $\phi(1020)$ are 
found to be strictly zero in our model since a $s\bar s$ state
clearly cannot decay into two $n\bar n$ states via a simple quark loop
mechanism alone; the non--vanishing experimental widths for
$\phi(1020)\to\pi\pi/\omega\pi$ indicate that additional diagrams
beyond lowest order in fact play a role for some decays even if mesons
with non--zero total angular momentum are involved in the process. For the
decays  of the $\phi(1680)$ and $\phi_3(1850)$ into kaon pairs, we
find at least plausible results by considering the quark loop
contribution; again, the sums of the $KK$ and $KK^*$ decay widths are
not sufficient to explain the total decay widths of these mesons.


\subsection{Decays of \boldmath$h_J$\unboldmath\/ and \boldmath$b
_J$\unboldmath\/ Mesons}\label{bhJ_SUBSECTION} 

The decay widths of the isoscalar $h_J$ and isovector $b_J$ mesons are
summarized in tab.~\ref{hb}. Note that the $n\bar n$ states
$h_1(1700)/b_1(1700)$ and $h_3(2050)/b_3(2050)$ are not listed by the
\PDG (see \cite{PDG2000}); however, their (degenerated) masses assumed
in the framework of the $^3P_0$ model in
ref. \cite{BarnesStrongDecays2} fit excellently to the numerical
results of our relativistic quark model (see
\cite{KollRicken,KollPhD}). Thus, we have decided to list 
their decay widths although no experimental data concerning their
partial widths exist so far.

The $\rho\pi$ decay modes of the $h_J$ mesons are not as dominant in
our calculation as they appear in the $^3P_0$ model. In fact, our
calculation yields a larger partial width if the first radial $\rho$
excitation is in the final state. Concerning the vanishing probability
of the decays into a $b_1(1235)\pi$ final state, we observe a
similarity between our approach and the $^3P_0$ model.
The comparison concerning the other $b_J$ decay modes show that the $\rho\rho$ final
state is strongly suppressed in our calculation while --- in the
framework of the $^3P_0$ model --- it plays a
significant role in the $b_1(1700)$ and $b_3(2050)$ decays. We refrain from a further
discussion of this sector as there are no reliable experimental data
to compare with so far.


\subsection{Decays of \boldmath$a _J$\unboldmath\/ Mesons}\label{aJ_SUBSECTION}

We review our results concerning the $a_J$ strong decays in
tab.~\ref{a}. The masses of these isovector mesons are quite well
decribed as the $a_0$ masses are only slightly overestimated while the
$a_J$ masses for $J\not =0$ are correctly calculated up to only a few
percent in our approach (see \cite{KollRicken}); note that this
statement also holds for the radial excitations in this sector.

\begin{figure}
\begin{center}
\input{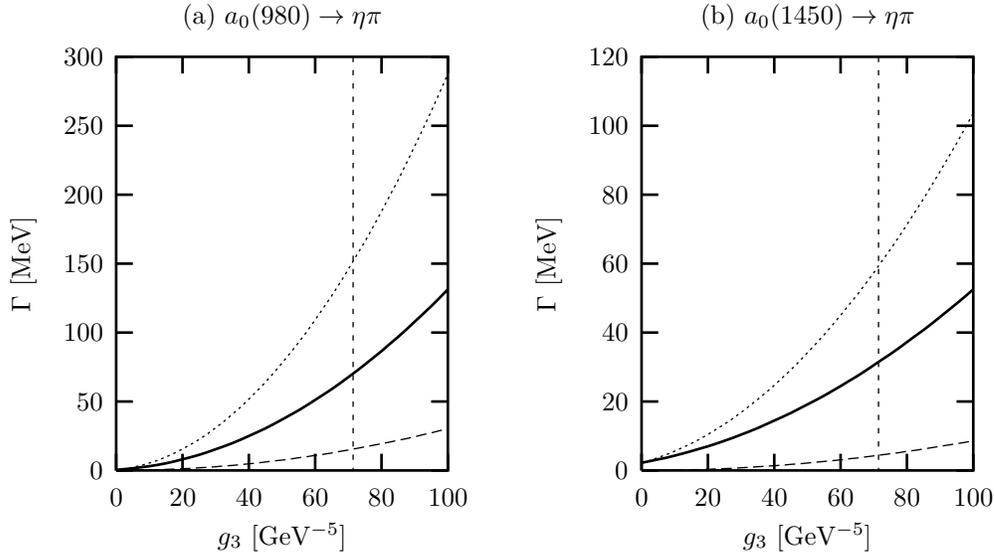}
\caption{{\small The widths of the decays (a) $a_0^*(980)\to \eta\pi$ and
(b) $a_0^*(1450)\to \eta\pi$ and their dependence on the coupling constant
$g^{(3)}_{\mbox{\scriptsize eff}}$ of 't~Hooft's three--body force due
to instanton effects; see text for more details. The fat solid line
({\bf ---}) indicates the total decays widths based on the
interference of OZI allowed diagrams (-- -- --) and OZI violating
diagrams due to the instanton interaction ($\cdots$); obviously, the
interference in this example is  destructive. Again, the vertical
dashed lines mark our choice for the 't~Hooft coupling
$g^{(3)}_{\mbox{\scriptsize eff}}$.}} 
\label{fig:a0_Decays} 
\end{center}
\end{figure}

Let us first focus on the strong two--body decays of the $a_0(980)$
and $a_0(1450)$ mesons. The relative strength of the $\eta\pi$ and $KK$ decay
modes are well reproduced for the $a_0$ ground state. For the
radial excitation, the pattern is again quite plausible: the decay widths
add up to $\approx 100$~MeV compared to only $\approx 10$~MeV in the
$^3P_0$ model which strongly underestimates the experimentally
determined total decay width. In fig.~\ref{fig:a0_Decays}, we show how
the instanton--induced six--quark interaction significantly lifts the
$\eta\pi$ partial width of the $a_0(980)$
and $a_0(1450)$ mesons; without this contribution (\ie for
$g^{(3)}_{\mbox{\scriptsize eff}}=0$), the quark loop diagram alone
would in fact give results of only a few MeV as predicted by the
$^3P_0$ model. We consider the interplay between the quark loop and
the instanton--induced contribution to these decay widths as an
impressing feature of our ansatz that includes diagrams beyond lowest
order. The interference of OZI allowed and OZI forbidden contributions
can clearly be studied in fig.~\ref{fig:a0_Decays}; obviously, the
Zweig rule violating processes are important for the understanding of
these decays. In this example, the relative sign between both parts
leads to a destructive interference finally yielding 
partial widths in the correct order of magnitude. This underlines our
introductory statement that the instanton--induced mechanism plays a
crucial role in the description of scalar meson decays.

The lack of diagrams induced by 't~Hooft instanton force is visible in
our results for the $a_1$ mesons. They are smaller by an order of
magnitude compared to the non--relativistic $^3P_0$ model; we thus
find again that the pure quark loop calculations underestimate the
partial widths in the presence of one or two pions in the final
state. However, the decays of the $a_2(1320)$ are plausibly described
in our relativistic approach compared to the $^3P_0$ model as well as
to experiment. The accuracy of our results is not satisfying;
nevertheless, the numbers in tab.~\ref{a} clearly show that an
approximative agreement with the experimental data for the $a_2(1320)$
partial widths can be achieved without any re--adjustment of our
fundamental model parameters.


\subsection{Decays of \boldmath$f _J$\unboldmath\/ Mesons}\label{fJ_SUBSECTION}

The strong decays of the isoscalar $f_J$ mesons are particularly
interesting as they might allow for an identification of the possible
non--$q\bar q$ nature of certain mesons in this sector. In
tabs.~\ref{f} and \ref{f2}, we summarize our results concerning the
$f_J$ mesons and compare them with experimental data and (rare) results of
the $^3P_0$ model. Note that the masses of these mesons are correctly
determined within our approach up to only a few percent (see
\cite{KollRicken,KollPhD}). 

\begin{figure}
\begin{center}
\input{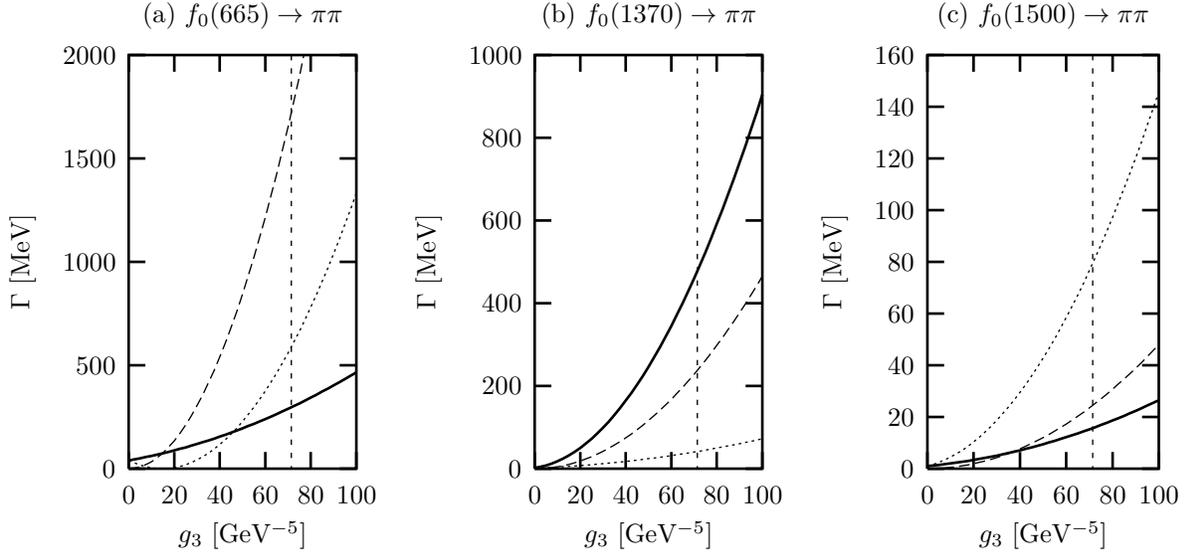}
\caption{{\small The widths of the decays $f_0\to \pi\pi$ and their
dependence on the coupling constant $g^{(3)}_{\mbox{\scriptsize eff}}$
of 't~Hooft's three--body force for (a) the ground state with
$M_{f_0}\approx 400\ldots 1200$~MeV$\approx 665$~MeV, (b) the first radial excitation with
$M_{f_0}\approx 1370$~MeV, and (c) the second radial excitation with
$M_{f_0}\approx 1500$~MeV; see text for more details. Again, the fat solid
line ({\bf ---}) indicates the total decays widths based on the
interference of OZI allowed diagrams (-- -- --) and OZI violating
diagrams due to the instanton interaction ($\cdots$). Here as in the other
pictures, the vertical dashed lines mark our choice for the 't~Hooft
coupling $g^{(3)}_{\mbox{\scriptsize eff}}$.}} 
\label{fig:f0_pi_pi} 
\end{center}
\end{figure}

Let us first consider the decays $f_0\to\pi\pi$ for the ground state
($M_{f_0}\approx 400\ldots 1200$~MeV$\approx 665$~MeV), the first
($M_{f_0}\approx 1370$~MeV) and the second ($M_{f_0}\approx 1500$~MeV)
radial excitation. We have plotted the partial decay widths with
respect to their dependence on the six--quark coupling
$g^{(3)}_{\mbox{\scriptsize eff}}$ of 't~Hooft's force in
fig.~\ref{fig:f0_pi_pi}. As all mesons in these decays have total
angular momentum $J=0$, the instanton--induced interaction provides an
additional contribution to the decay widths beyond the lowest order
quark loop diagrams. Moreover, we observe Zweig rule violating
amplitudes in these decays as the $f_0$ mesons are isoscalar (see
section \ref{CalculationOfTheDecayaWidth} for comparison). The widths
based on OZI allowed processes and OZI forbidden processes are
indicated in fig.~\ref{fig:f0_pi_pi}; the fat solid line in these
plots denote the total partial width for $f_0\to\pi\pi$ produced by
the interference of the underlying mechanisms.  

At this point, we want to comment on the frequently discussed
considerations of the scalar nonet in standard (mainly
nonrelativistic) quark models: Due to the suppression of the decay
mode $f_0(1500)\rightarrow K \bar K$, the $f_0(1500)$ should be
assigned to a dominantly $\bar n n$-state rather than to a dominantly
$\bar s s$-state. Then this resonance may be viewed as the isoscalar
partner of the $a_0(1450)$ and $K_0^*(1430)$. Whereas the observed
widths for the latter resonances ($\sim 300$ MeV) can be well
reproduced in the standard quark models, the same models yield
$\Gamma(f_0(1500)) > 500$ MeV in clear contradiction to
experiment. From this point of view the experimental width of the
$f_0(1500)$ indeed seems to be ``unnaturally small''. However, these
arguments are completely wrong if Zweig rule violating forces are
taken into account; merely, the conventional conclusion concerning the flavour
contents of resonances, which may be roughly formulated as ``dominant $\pi\pi$ decay
mode indicates large nonstrange contents in the decaying resonance'' and
``dominant $K\bar K$ decay mode indicates large strange contents in the
decaying resonance'', is certainly not true for mesons that can decay via the
instanton induced six-quark interaction via a Zweig rule violating
process.

In fig.~\ref{fig:f0_pi_pi}(a), the decay $f_0(400\ldots
1200)\to\pi\pi$ is plotted; obviously, the interference between OZI
allowed processes and OZI forbidden processes is destructive. The
Zweig rule violating width is huge at $g^{(3)}_{\mbox{\scriptsize
eff}}=71.4$~GeV$^{-5}$; due to the destructive interference
of the different amplitudes, the resulting partial width is however
reduced to only 297~MeV which is quite realistic for this decay
mode. Note that the pure quark loop contribution in this example is
only $\approx 40$~MeV which would clearly underestimate the $\pi\pi$
width of the $\sigma$ meson; again, we thus see the interesting
influence of the additional instanton--induced contribution to the
partial width. The analogue plots for the decays
$f_0(1370)\to\pi\pi$ and $f_0(1500)\to\pi\pi$ are given in
figs.~\ref{fig:f0_pi_pi}(b) and \ref{fig:f0_pi_pi}(c). For the
$f_0(1370)$ decay, we observe a constructive interference of the OZI
allowed and OZI forbidden amplitudes lifting the partial width to
477~MeV which is comparable to the $^3P_0$ model with 271~MeV (see
\cite{BarnesStrongDecays1}). However, the experimental width given in
ref. \cite{Abele} is significantly smaller, namely
$\Gamma_{\mbox{\small exp}}=21.7\pm9.9$~MeV. In the {\sc Crystal
Barrel} experiment presented in this publication, the $f_0(1370)$
dominantly decays into its $\sigma\sigma$ mode. Unfortunately, we
cannot compute this decay for a more detailed comparison due to
kinematical reasons\footnote{As in the decay $\phi(1020)\to K K$, the
exact masses of the outgoing particles in $f_0(1370)\to\sigma\sigma$
are slightly larger by some MeV than the mass of the decaying
particle; in section \ref{omegaphiJ_SUBSECTION}, we explain why we
refrain from a parameter re--adjustment in such a situation.}; it is
therefore difficult to explain this discrepancy of the $\pi\pi$
partial width between theory and experiment from our point of view. In
ref. \cite{Abele}, we also find an experimental value for the partial
width of the decay $f_0(1500)\to\pi\pi$ which reads
$\Gamma_{\mbox{\small exp}}=44.1\pm 15.3$~MeV. Again, the experimental
value is one order of magnitude smaller than the several hundred MeV
that we found for the $\pi\pi$ decay modes of the first two states in the $f_0$
spectrum. Surprisingly, our result for this decay width is 15.7~MeV
which is in rough agreement with the experiment due to the destructive
interference of the various amplitudes that contribute to this
process. Without 't~Hooft's six--quark interaction, the pure quark
loop result for the partial width would be $\approx 1$~MeV which would
have clearly underestimated the experimental data. This observation
again underlines the importance of the unique feature of our approach,
namely the generic inclusion of instanton--induced vertices beyond
lowest order for the determination of the strong two--body decay
widths. For further results of the $^3P_0$ model for the
$f_0$ and the $f_1$ mesons, we refer to the  recent preprint of Barnes {\sl et al.} (see \cite{Barnes:2002mu}).  

\begin{figure}
\begin{center}
\input{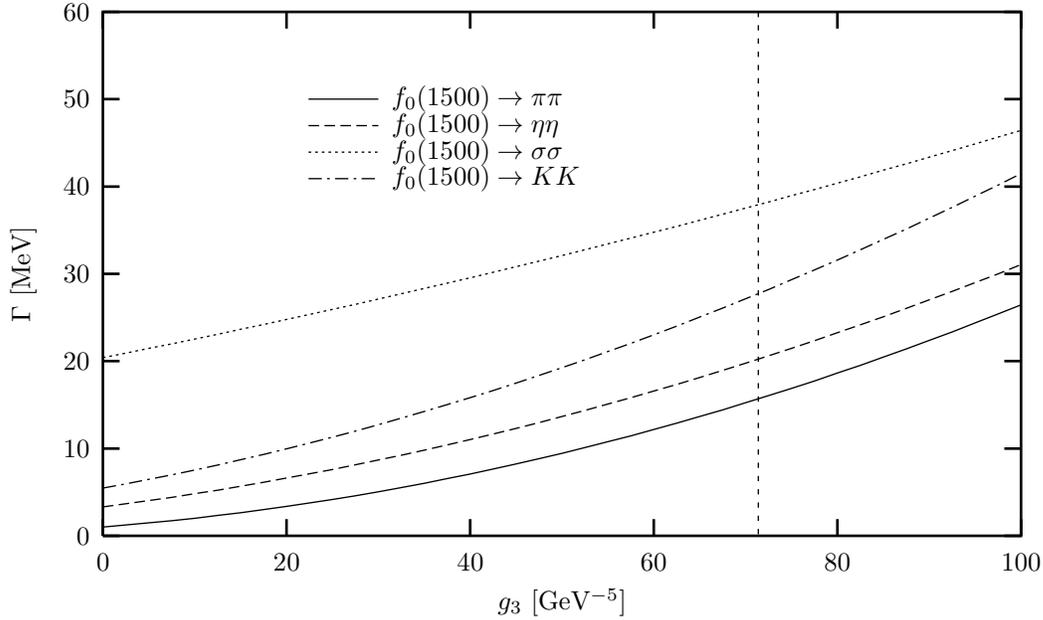}
\caption{{\small The widths of the decays $f_0(1500)\to \pi\pi$,
$f_0(1500)\to \eta\eta$, $f_0(1500)\to \sigma\sigma$ and $f_0(1500)\to KK$,  and their
dependence on the coupling constant $g^{(3)}_{\mbox{\scriptsize eff}}$
of 't~Hooft's three--body force; see text for more details. The
vertical dashed lines mark our choice for the 't~Hooft coupling
$g^{(3)}_{\mbox{\scriptsize eff}}$.}}  
\label{fig:f0_1500_Decays} 
\end{center}
\end{figure}

Let us now turn to a discussion of the other $f_0$ decay modes which
we have not mentioned so far. In tab.~\ref{f}, we present numerous
other results of our calculations with respect to $\eta\eta$,
$\sigma\sigma$, $K K$ and other final states. In most cases, we find
the correct order of magnitude for the widths although the results are
sometimes significantly off the error bars. In fig.~\ref{fig:f0_1500_Decays}, we
plot the partial widths for the decays
$f_0(1500)\to\pi\pi/\eta\eta/\sigma\sigma/ KK$ with respect to the
strength of the instanton--induced six quark interaction. The results at
$g^{(3)}_{\mbox{\scriptsize eff}}=71.4$~GeV$^{-5}$ in general tend to
be too large compared with the experimental widths presented in
ref. \cite{Abele}. This fact might hint at a possible re--adjustment
of the six--quark coupling constant $g^{(3)}_{\mbox{\scriptsize eff}}$
at these data; as we have noted before, we have hitherto fixed this
constant at the partial width of the decay $K_0^*\to K\pi$ where it
might have been adjusted to a slightly too large numerical value. We
shall note at this point that the experimental data in this sector are
from one experiment only (see \cite{Abele}) and that there are no
other theoretical calculations such that a detailed comparison might
be premature. However, our results seem to be not completely
unplausible such that we still consider our approach as a reliable
basis for future studies in this field.  

Another point to mention here is the flavour structure of the $f_J$
mesons with $J\not =0$. In general, the ground state is purely $n\bar
n$ while the next radial excitation is a $s\bar s$ state; note that
--- in contrast to the $f_0$ mesons --- there is no flavour mixing for
these mesons as 't~Hooft's force only acts on mesons with $J=0$. This
is the reason why the $\pi\pi$ widths of the first excitations of the
$f_1$, $f_2$ and $f_4$ mesons are zero since the quark loop mechanism
alone cannot provide a $s\bar s\to n\bar n+ n\bar n$
transition.\footnote{For the same reason, we find vanishing decay
widths for the decays $f_4(2220)\to\omega\omega$ and $f_4(2220)\to a_2\pi$; see
tab.~\ref{f2} for comparison.} The $\pi\pi$ widths of the related
ground states are in general too low by orders of magnitudes; as we
have already concluded before, the instantaneous Bethe--Salpeter
ansatz might not be adequate for the deeply bound pion such that the
resulting meson--quark--anti-quark vertex functions are not
appropriate for further calculations even in a relativistic
framework. The other decay modes of the $f_J$ mesons ($J\not =0$) are in
acceptable agreement with the experimental data; in particular the
$f_2(1270)$ and $f_2'(1525)$ decays into $KK$ and $\eta\eta$ final
states are excellently described.


\subsection{Decays of \boldmath$K _J$\unboldmath\/ Mesons}\label{KJ_SUBSECTION}

In tab.~\ref{K}, we summarize our calculations of partial widths for
strongly decaying $K_J$ mesons. Let us note that the $K_J$ states for
$J\not =0$ are twice degenerated for each mass due to spins $S=0$ and
$S=1$ which are not separated by the interquark forces in our model;
nevertheless, the number of states in the spectrum is correct (see
\cite{KollRicken}). 

The $K(1460)$ is the first radial excitation of the pseudoscalar kaon
in our model. In our calculation, the $K\rho$ mode is quite strong
compared to the other final states; this seems to be not unrealistic
compared to experiment. Note hereby that the nearly
vanishing width for $K(1460)\to K_0^*(1430)\pi$ is found due to the
destructive interference between the approximately equal amplitudes
from the quark loop diagram and the instanton--induced six quark
vertex contributing to this particular process.

The decays $K_1(1270)\to K\rho/K^*\pi/ K\omega$ are excellently
described in our model; here, the $^3P_0$ model clearly overestimates
all widths in this sector. This impressing agreement is exceptional
insofar as other kaonic decay modes often suffer from the unrealistic
smallness of the calculated partial widths. Indeed, for the next $K_1$ excitation,
our results are less well in agreement with the experiment. Moreover,
the $K_2(1580)$ and $K_2(1770)$ decays seem to be clearly
underestimated in our approach; the experimental data are however poor
and no comparison with the non--relativistic model is possible so
far.


\subsection{Decays of \boldmath$K^* _J$\unboldmath\/ Mesons}\label{KstarJ_SUBSECTION}

As we have already discussing in section \ref{Fixing_g3}, the value of
the strength parameter for 't~Hooft's six quark interaction is fixed
at the experimental decay width $\Gamma_{\mbox{\small exp}}=274\pm
37$~MeV of the decay $K_0^*(1430)\to K\pi$. In
fig.~\ref{fig:Fixing_g3}, we present the related plots not only for
the ground state but also for the first radial excitation
$K_0^*(1950)$ which is also very well described by our choice
$g^{(3)}_{\mbox{\scriptsize eff}}=71.4$~GeV$^{-5}$. Thus, the decay
widths for  $K_0^*\to K\pi$ presented in tab.~\ref{Kstar} are clearly
in good agreement with the experimental data since we have adjusted
our last free parameter $g^{(3)}_{\mbox{\scriptsize eff}}$ in this
sector. The other decays in this final table  involve mesons with
$J\not =0$; they proceed exclusively via quark loop diagrams and are
thus not affected by our parameter choice. 

It is very interesting to consider the decays $K_J^*\to K\pi$
with $J=1,2,3,4,5$; let us first restrict to ground state $K_J^*$
mesons. For $J=1$, we find an excellent agreement with the experimental
width $\Gamma_{\mbox{\small exp}}\approx 50$~MeV. The experimental
partial widths for $K\pi$ final states decrease characteristically for
increasing total angular momenta $J$. A comparison with the formula
given in eq. (\ref{DecayWidthFormula}) shows that this decrease can be
approximately traced back to the influence of the phase space factor
$\frac{1}{2J_1+1}\cdot\frac{k}{M^2_1}$ such that the matrix element
itself should be more or less the same for all $K_J^*\to K\pi$ decays
($J=1,\ldots, 5$). Unfortunately, this is not correctly described by
our model such that our calculation increasingly fails to describe
correctly the experimental partial widths for higher angular momenta
$J$. We have already noted that 
pionic final states might be hard to describe in an instantaneous
framework; here, we find another example for this
conjecture. 

For the other decay modes in this sector, we observe generally a quite
poor agreement with experiment as well as with the results of the
$^3P_0$ model. For particular decay channels, our calculation yields
more realistic numbers for the partial widths than the
non--relativistic model (\eg for $K_2^*(1430)\to K\eta$) but, in
general, we observe a decreasing accuracy of our results for higher
$K_J^*$ excitations and larger total angular momenta. Note however that the calculated large widths for the decays
$K_3^*(1780)\to K^*(892)\rho$ and $K_3^*(1780)\to K^*(892)\omega$ lead at least to a qualitative understanding of the total decay width in the framework of our model.


\subsection{Decays of \boldmath$\eta _J$\unboldmath\/ Mesons}\label{etaJ_SUBSECTION}

In tab.~\ref{eta}, we present our results concerning the strong decays
of the $\eta'(958)$, $\eta(1440)$ and $\eta_2(1645)$ mesons. Their
masses are well described in our relativistic quark model; the
well--known flavour mixing of the pseudoscalar $\eta$ mesons is
induced by 't~Hooft instanton interaction and has been extensively
discussed in refs. \cite{KollRicken,KollPhD}. We should note that our
quark model does not support the hypothesis of the existence of the
$\eta(1295)$ meson as a pure $q\bar q$ state; it is interesting that
in fact no evidence for this state has been found by the {\sc Crystal
Barrel} group (see \cite{Suh,Reinnarth,Suh2,Suh3}).

\begin{figure}
\begin{center}
\input{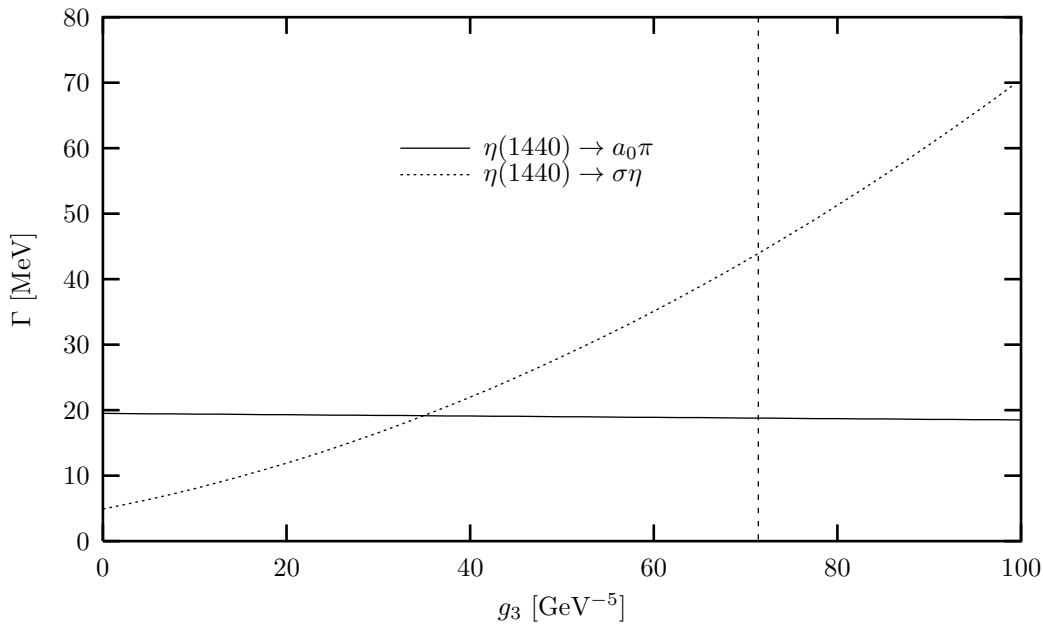}
\caption{{\small The widths of the decays $\eta(1440)\to
a_0^*(980)\pi$ and $\eta(1440)\to\sigma\eta$ and their dependence on
the coupling constant $g^{(3)}_{\mbox{\scriptsize eff}}$ of 't~Hooft's
three--body force due to instanton effects; see text for more
details. Note that the vertical dashed line denotes our choice for the
't~Hooft coupling $g^{(3)}_{\mbox{\scriptsize eff}}$.}} 
\label{fig:eta_Decays} 
\end{center}
\end{figure}

The vanishing decay width of the process $\eta'\to\rho\pi$ is in
agreement with experiment. Some preliminary partial decay widths of
the $\eta(1440)$ meson are quoted in ref. \cite{Suh}; they are given
in tab.~\ref{eta}. Our results are in acceptable (and partly
excellent) agreement with these experimental data; note that this
statement refers to decays that are described by quark loop diagrams
only (such as the $KK^*$ final state) as well as to decays in which
additional instanton--induced vertices have been considered (such as the
$a_0\pi$ and $\sigma\eta$ final states). Concerning the latter decays,
the {\sc Crystal Barrel} group recently found the experimental decay
width ratio $R:=\Gamma_{\mbox{\small exp}}^{a_0\pi}/\Gamma_{\mbox{\small
exp}}^{\sigma\eta}=0.62\pm 0.14$ (see \cite{Reinnarth}). In our
approach, we find $R=0.43$ which is in reasonable agreement with this
analysis; note that $R=0.4\pm 0.2$ has earlier been found in
ref. \cite{Abele} (for more references, see \cite{Reinnarth}). It is
interesting to look at fig.~\ref{fig:eta_Decays} for a more detailed
discussion; in this plot, we present the dependence of the decay
widths $\eta(1440)\to a_0\pi$ and $\eta(1440)\to\sigma\eta$ on the
strength of 't~Hooft's instanton force. Since the $a_0\pi$ partial
width remains more or less constant while the $\sigma\eta$ width is
monotonically increasing, a sufficiently large coupling constant
$g^{(3)}_{\mbox{\scriptsize eff}}$ is needed in order to compute a
ratio $R<1$; however, we find again that a slightly smaller value than
$g^{(3)}_{\mbox{\scriptsize eff}}=71.4$~GeV$^{-5}$ would provide a
more exact ratio compared to the experiment quoted in
ref. \cite{Reinnarth}. Finally, the $\eta_2(1645)$ decay widths are
smaller than the values obtained in the framework of the $^3P_0$
model; a more detailed comparison is prevented by the poor accuracy of
the experimental data.


\section{Summary}
\label{Summary}

In this paper, we have studied strong two--body decays of light
mesons. Our approach is based on a relativistic quark model which
describes the complete meson spectrum by means of the Bethe--Salpeter
equation in its instantaneous approximation. The quark--anti-quark
interaction is parameterized by a confinement potential with an
appropriate spinorial structure in Dirac space. Furthermore, we adopt
't~Hooft's instanton--induced interaction that provides a conving
picture for the well--known mass splittings and flavour mixing effects
in the pseudoscalar and scalar meson sector. As we have shown in
refs. \cite{KollRicken,RickenKoll}, this formally covariant ansatz
leads to an excellent description of the mass spectrum of the light
mesons. Furthermore, we have investigated several electroweak decay
processes involving light mesons with good results compared to the
experimental data; a complete overview on this approach for the
intense study of light mesons can be found in ref. \cite{KollPhD}. 

For the calculation of the strong decay widths, we have applied the
Mandelstam formalism for the derivation of the related matrix
element. Hereby, we have included not only the lowest order quark loop
diagrams that contribute to all possible decay modes; moreover, we
implemented the three--body interaction of 't~Hooft's
instanton--induced force that acts on mesons with vanishing total
angular momentum. We have thus found that an interference between
quark loop diagrams and amplitudes originating in the instanton
interaction occurs if all mesons in the decay $\123$ have angular
momentum $J_i=0$ ($i=1,2,3$). We shall add that 't~Hooft's six quark
vertices also provide Zweig rule violating amplitudes contributing to
the partial width if and only if at least one of
the mesons in the decay process is of isoscalar type. Summarizing, we
have calculated the quark loop for each decay and, additionally, the
instanton--induced contribution if only mesons with $J_i=0$
participate the process under consideration; eventually, OZI forbidden
amplitudes might occur due to the latter mechanism if $\eta,\eta',\ldots$ or
$f_0,f_0', \ldots$ mesons are involved in the particular decay.

We have compared our numerical results for the numerous partial decay
widths with the experimental data, if possible (see \cite{PDG2000} and
\cite{Abele,Suh,Barberis,Bugg,Reinnarth}); in some sectors, these
data are quite poor such that more reliable experimental results for the decay
width would be extremely helpful to clarify some of the puzzling
issues in this field. Moreover, we refered to the so--called $^3P_0$
model first proposed by A.~Le~Yaouanc \etal (see
\cite{LeYaouanc1,LeYaouanc2,LeYaouanc3}) which is a non--relativistic
approach using quark lines and assuming the creation of an additional
quark--anti-quark pair with vacuum quantum numbers $J^{\pi c}=0^{++}$;
the numerical results obtained in this framework were quoted from
refs. \cite{BarnesStrongDecays1,BarnesStrongDecays2,BonnazSilvestreBrac}. 

Before we summarize our numerical results for the partial widths, we
shall note that our approach --- similar to the $^3P_0$ model --- does
not take into account any final state interaction. The neglect of the
related effects is clearly a shortcoming of our ansatz which is,
however, hard to overcome in a quark model like the one presented in
this publication. We have thus restricted our model to the ``impulse
approximated'' approach although final state interactions are expected to
significantly modify the partial widths. 

Let us first focus on decays in which only the pure quark loop
diagrams contribute to the partial widths. We found that we achieve a
description of the strong meson decays that is in parts comparable to
the $^3P_0$ model. However, we also noted that we sometimes
underestimate the partial widths by one order of magnitude (or even
more) compared to
the experiment. This flaw is mostly observed for decays in which one
or two pions are in the final state; it might be related to the fact
that the underlying instantaneous approximation of the Bethe--Salpeter
equation is not suitable for deeply bound states such as the pion
ground state. Furthermore, decays of mesons with large angular
momentum and decays of highly excited mesons are in general not so
well described by the quark loop mechanism alone. It remains an open
question whether additional effects due to the final state interaction
could modify this observation.

For decays in which instanton--induced vertices beyond the lowest
order quark loop diagrams are included, we find that the additional six
quark interaction has a strong impact on the numerical results. These
decays involve only scalar and pseudoscalar mesons; the related
coupling strength $g^{(3)}_{\mbox{\scriptsize eff}}=71.4$~GeV$^{-5}$
has been fixed to the decay $K_0^*\to K \pi$ with $\Gamma_{\mbox{\small
exp}}=274\pm 37$~MeV. In this example, 't~Hooft's three--body force
lifts the partial decay width from a few MeV from the pure quark loop
mechanism up to several hundred MeV in excellent agreement with the
experiment. Here, the interference between both mechanisms was
constructive; we have however discussed several other decay modes in
which the interference was destructive. The interference between quark
loop amplitudes and instanton--induced amplitudes can be translated
into an interference between OZI allowed and OZI forbidden amplitudes
if isoscalar mesons are involved. In these cases, we discussed
numerous examples in which the (constructive or destructive)
interference between both contributions led to very realistic results
for the partial decay widths. In general, we found that the additional
instanton--induced mechanism significantly modified the result of the
pure quark loop calculation. Hereby, we discussed that the conventional conclusion concerning the flavour contents of
resonances, which may be roughly formulated as ``dominant $\pi\pi$ decay
mode indicates large nonstrange contents in the decaying resonance'' and
``dominant $K\bar K$ decay mode indicates large strange contents in the
decaying resonance'', is not true for mesons that can decay via the Zweig rule violating part of the 
instanton induced six-quark interaction and may be
misleading. Summarizing, we presented impressing examples for the
importance of this contributions beyond lowest order and concluded
that, apart from the quark loop diagrams, the inclusion of additional
effects at least for scalar and pseudoscalar meson decays seems to be
mandatory. 

For future studies, it might be helpful to consider effects that are
not included in our approach such as final state interactions or other
mechanisms beyond lowest order or instanton--induced vertices;
moreover, more experimental data are clearly needed for a better
assessment of the various theoretical models for this problem. The
study of strong mesons decays thus remains a challenge for theory and
experiment as well. However, we believe that the present publication
provides a reliable framework for a detailed investigation of strong
meson decays as it is based on a completely covariant approach and
includes realistic mechanisms for the description of the decay
processes.


\small

\section*{Acknowledgements}

We have profited very much from discussions with H.--R.~Petry,
J.--S.~Suh, B.~Pick, J.~Reinnarth, A.~V.~Sarantsev and E.~Shuryak to
whom we wish to express our gratitude. Furthermore, we thank
V.~Hellmann for useful numerical checks and careful reading of the
manuscript. Financial support of the {\sc Deutsche
Forschungsgemeinschaft} is gratefully acknowledged.


\normalsize


\newpage

\begin{appendix}

\section{Flavour Factors and Charge Multiplicities}
\label{app:FlavourFactors}

For the calculation of the general (\ie charge independent) widths of
decays like $\rho\to\pi\pi$ instead of a special process with definite
charges like $\rho^+\to\pi^+\pi^0$, it is well known that one only
needs to sum over all possible final states and to average over all
possible initial states. Using isospin symmetry, this operation links
the special decay width $\Gamma _{\123}^{\mbox{\scriptsize special}}$
with the general decay width $  \Gamma _{\123}^{\mbox{\scriptsize
general}} $ by the use of a Clebsch--Gordan coefficient $C:=\langle
I_2 m_2\: I_3 m_3 | I_1 m_1 \rangle$; here, $I_k$ denotes the isospin
with its third component $m_k$ ($k=1,2,3$). For bosonic particles in
the final states such as mesons, one has to add a symmetry factor
$\frac 1 2$ if both outgoing particles are identical; note that this
factor of course must not be multiplied for a $K\bar K$ final
state. Thus, the expression 

\equ{
\Gamma _{\123}^{\mbox{\scriptsize general}} = \Gamma
_{\123}^{\mbox{\scriptsize special}} \:\:\frac{1 - \frac 1 2\delta 
_{{\cal M}_2{\cal M}_3}}{\Big|\:\big\langle\: I_2
m_2\:\: I_3 m_3 \:\big|\: I_1 m_1 \:\big\rangle\:\Big|^2}
}

yields the correct general decay width $
\Gamma _{\123}^{\mbox{\scriptsize general}} $  although in fact only
the particular width $\Gamma _{\123}^{\mbox{\scriptsize special}}$ 
is numerically  computed (see also \cite{BarnesStrongDecays1}). Note that the
evaluation of the flavour trace in the decay matrix element actually
yields a factor $f$ which clearly depends on the particular charge
distribution in the special decay. As examples, we quote
$f^2(K^{*+}\to K^0 \pi^+)=1$ and $f^2(K^{*+}\to K^+ \pi^0)=\frac 1 2$
for the decay of an isodublet meson into an isodublet meson and an
isovector meson. Here, the Clebsch--Gordan coefficients $C^2(K^{*+}\to
K^0 \pi^+)=\frac 1 3$ and $C^2(K^{*+}\to K^+ \pi^0)=\frac 2 3$ in isospin 
space are different for both decay types such that the fraction ${\cal
F}_{K^*\to K\pi}:= f^2_{K^*\to K\pi}/C_{K^*\to K\pi}^2=3$ is
constant. For the sake of clarity, we quote all possible flavour
weight factors in tab.~\ref{FlavourFactors}; we refer to ref.
\cite{BarnesStrongDecays1} for further comments.


\section{The Instanton--Induced 't~Hooft Interaction}
\label{app:tHooft}

Following G.~'t~Hooft's seminal ideas on instantons in QCD first published
in ref. \cite{tHooft}, M.~A.~Shifman, A.~I.~Vainshtein and
V.~I.~Zakharov have derived the Lagrangian $\Delta {\cal
L}^{\mbox{\scriptsize eff}}$ describing the contribution of
instanton--anti-instanton configurations to the effective Lagrangian
for light quarks (see \cite{ShifmanVainshteinZakharov}). Their result is discussed in
ref. \cite{ResagMünzSecondPaper} with respect to the relativistic quark
model for mesons which provides the basis for the present
contribution. 

Omitting the confining interaction for a moment, we can write down the
following effective Lagrangian mimicking the QCD dynamics in our
model:

\equ{
\label{tHooftLagrangian}
{\cal L}_{\mbox{\tiny QCD}}^{\mbox{\scriptsize eff}} &=& {\cal
L}_{\mbox{\tiny QCD}}^0 + \Delta {\cal
L}^{\mbox{\scriptsize eff}} = k + \sum _{f=1}^3 \::\: \bar\Psi_f
\left(i\myslash{\partial}-m_f\right)\Psi_f\::\: + {\cal L}^{(2)} +
{\cal L}^{(3)}  \quad;
}
here, ${\cal L}_{\mbox{\tiny QCD}}^0=\sum _{f=1}^3 \bar\Psi_f
(i\myslash{\partial}-m_f^0)\Psi_f$ is the free quark
Lagrangian with current quark masses $m_f^0$ (for the flavour index,
$f=\{1,2,3\}=\{u,d,s\}$ holds). In ${\cal
L}_{\mbox{\tiny QCD}}^{\mbox{\scriptsize eff}}$, the masses
$m_f=m_f^0+\Delta m_f$ are constituent quark masses with a
characteristic contribution $\Delta m_f$ to the effective mass due to
instanton effects (see \cite{ResagMünzSecondPaper}). The constant $k$
denotes the vacuum energy density and is inessential for our
present considerations. The contributions ${\cal L}^{(2)}$ and ${\cal
L}^{(3)}$ represent the two--body and the three--body interactions,
respectively; we will briefly discuss these terms in the following


\subsection{The Two--Body Interaction}
\label{tHooft2Body}

As it has been discussed in ref. \cite{ResagMünzSecondPaper}, the
relativistic quark model used in this publication incorporates a
particular residual interaction besides the global confining
interaction. It is described by the two--body term in 't~Hooft's
instanton--induced force and reads explicitly

\equ{
\label{L2}
{\cal L}^{(2)} = g_{\mbox{\scriptsize eff}}^{(2)} (i) \: \frac{3}{16}\:
\Big[\::\: \bar\Psi_k\bar\Psi_l\:\:
\big( \Id\otimes\Id + \gamma_5\otimes \gamma_5\big)\:\epsilon_{ikl}\epsilon_{imn}\:\big(2{\cal
P}^C_{\bar 3}+ {\cal P}^C_6\big)\:\: \Psi_m\Psi_n \::\:\Big]\:.
}

Here, a summation over the flavour indices $i,k,l,m,n=\{1,2,3\}=\{u,d,s\}$ is
understood. Note that ${\cal P}^C_{\bar 3}$ denotes the projector onto
colour anti--triplet and ${\cal P}^C_6$ denotes the projector onto
colour sextet.

This instanton--induced two--body interaction only acts on $q\bar q$
bound states with total angular momentum $J=0$, \ie for pseudoscalar
and scalar mesons. Its effects on the mass spectrum of these states
are highly interesting; they have been intensively discussed in
refs. \cite{KollRicken,RickenKoll}. In our calculations for light mesons, we assume
SU(2) flavour invariance and define for practical reasons the coupling
constants $g:=\frac 3 8 g_{\mbox{\scriptsize eff}}^{(2)}(s)$ and $g':=\frac 3
8 g_{\mbox{\scriptsize eff}}^{(2)}(n)$ where $s$ and $n$ denotes ``strange''
and ``non--strange'' quarks, respectively. Note that this interaction
also provides a realistic mechanism for the well--known flavour mixing
effect in the isoscalar states. In this appendix, we refrain from
a discussion of this particular aspect and also of further implications of the
use of this residual interaction in our quark model; instead, we refer to
former publications for more details (see \cite{KollRicken,RickenKoll} and references therein).


\subsection{The Three--Body Interaction}
\label{tHooft3Body}

The three--body term ${\cal L}^{(3)}$ in eq. (\ref{tHooftLagrangian})
does not influence the spectroscopical results for $q\bar q$ mesons;
interestingly, it does neither contribute to baryons described by
colourless $qqq$ bound states (see \cite{Löring2}). However, it has
been shown in refs. \cite{RitterPaper,RitterDiplom,RickenPhD} that the
three--body term of 't~Hooft's interaction generates additional
contributions to strong two--body decays beyond the pure quark loop
part if all mesons participating the process have vanishing angular
momenta, \ie $J_1=J_2=J_3=0$. Written out explicitly, the term reads

\equ{
\label{L3}
{\cal L}^{(3)} = g_{\mbox{\scriptsize eff}}^{(3)}  \: \frac{27}{80}
\:\Big[\::\:\bar\Psi\bar\Psi\bar\Psi\:\:
\big(\Id\otimes\Id\otimes\Id+(\gamma_5\otimes\gamma_5\otimes\Id +
\mbox{cycl. perm.})\big) {\cal P}^F_1\big(2{\cal
P}^C_{10}+ 5{\cal P}^C_8\big)\:\: \Psi\Psi\Psi \::\:\Big]
}

where ${\cal P}^F_1$ is a projector onto a three--particle flavour
singlet state and ${\cal P}^C_{10}$ (${\cal P}^C_8$) is a projector onto
colour decuplet (octet). Due to the special Dirac structure of this
interaction, it can be written more convenient by using the Weyl
representation of the Dirac spinors, \ie 

\equ{
\Psi(x) = \left(\begin{array}{c}\xi(x)\\ \eta(x)\end{array}\right)
}

and $\gamma _5$ is diagonal in this representation. Then we can
express the three--body term of 't~Hooft's instanton--induced
interaction according to

\equ{
{\cal L}^{(3)} = g_{\mbox{\scriptsize eff}}^{(3)}  \: \frac{27}{30}
\:\Big[\::\:\eta^\dagger\:\eta^\dagger\:\eta^\dagger\:\:
{\cal O}^{FSC}\:\: \xi\:\xi\:\xi \::\:\Big]
\:+\:\big(\:\eta\:\longleftrightarrow\:\xi\:\big) \quad. 
}

Here, the operator ${\cal O}^{FSC}$ acts in flavour, spin and colour
space and reads explicitly 

\equ{
\label{O_FSC}
{\cal O}^{FSC} &:= & 2\:{\cal P}^F_1 \otimes{\cal P}^S_4 \otimes{\cal
P}^C_{10}  \:+\: 5\:{\cal P}^F_1 \otimes{\cal P}^S_2 \otimes{\cal
P}^C_8  
}

where ${\cal P}^S_{2}$ (${\cal P}^S_4$) is a projector onto
spin dublet (quadruplet). Note that the overall flavour projector
${\cal P}^F_1$ in the operator ${\cal O}^{FSC}$ leads to a
violation of the phenomenological OZI rule if one of the particles in the
decay process is an isoscalar meson; this point is briefly
discussed in section \ref{OZIViolation}.

\cleardoublepage

\section{Tables and Figures}

\subsection{Tables}

\begin{table}[b!]
\begin{center}
  \begin{tabular}{cclc}
\multicolumn{3}{c}{Parameter} &  This work\\
& & &  \\
\hline 
\hline 
& & &  \\
& $g$ &[GeV${}^{-2}$]             & $\phantom{-}$1.62  \\
{\sl 't~Hooft} &$g'$ &[GeV${}^{-2}$]   & $\phantom{-}$1.35  \\ 
{\sl interaction}& $g^{(3)}_{\mbox{\scriptsize eff}}$ &[GeV${}^{-5}$]    & $\phantom{-}$71.4  \\
& $\Lambda _{\mbox{\scriptsize
III}}$ &[fm] & $\phantom{-}$0.42\\  
 && &  \\
{\sl Constituent}& $m_n$& [MeV] & $\phantom{-}$ 380 \\
{\sl quark masses}&$m_s$& [MeV] & $\phantom{-}$ 550 \\
 & & & \\
 {\sl Confinement}& $a_c$& [MeV]    & --1135 \\
 {\sl parameters}& $b_c$ &[MeV/fm]  & $\phantom{-}$1300 \\
& &&  \\
{\sl Spin structure} &\multicolumn{2}{c}{${\mit\Gamma\otimes\Gamma}$}
 & 
$\frac 1 2 (\Id\otimes\Id - \gamma _\mu\otimes\gamma ^\mu - \gamma
^5\otimes\gamma ^5)$  \\
  \end{tabular}
\caption{The parameters of the confinement force, the 't
Hooft interaction and the constituent quark masses in this work; the
particular values of this parameter set correspond to model \B first
presented and dicussed in refs. \cite{KollRicken,RickenKoll}. Note
that the three--body 't~Hooft coupling 
$g^{(3)}_{\mbox{\scriptsize eff}}$ appears in this work for the first
time; it has been adjusted to the experimental width of the decay
$K_0^*\to K\pi$.}
\label{tab:Parameters}
\end{center}
\end{table}

\begin{table}[b!]
\begin{center}
\begin{tabular}{ccc@{\qquad}lc}
\quad$\: I_1\:$ \quad& $I_2$ & $I_2$ & Example\quad & \quad${\cal
  F}_{\123}$\quad\quad\quad \\
  & & & & \\
\hline
\hline
  & & & & \\
0  		& 0& 0 				& $f_2\to\eta\eta\,^\dagger$	& 1     \\[2ex]
$\frac 1 2$  	& 0& $\frac 1 2$ 		& $K^*\to K\eta'$	& 1     \\[2ex]
1  		& 0& 1 				& $a_2\to\pi\eta$	& 1     \\[2ex]
0  		&$\frac 1 2$ &$\frac 1 2$  	& $f_0\to K\bar K$	& 2     \\[2ex]
1  		&$\frac 1 2$ &$\frac 1 2$  	& $a_0\to K\bar K$	& 1     \\[2ex]
$\frac 1 2$  	&$\frac 1 2$ & 1 		& $K_2^*\to K\pi$	& 3     \\[2ex]
0  		&1 & 1 				& $f_0\to\pi\pi\,^\dagger$	& 3/2   \\[2ex]
1  		&1 & 1 				& $\rho\to\pi\pi\,^\dagger$	& 1     
\end{tabular}
\caption{Flavour weight factors for the
various isospin channels of the strong two--body decay $\123$; see
text for details. Note
that for the examples marked with $^\dagger$ an additional factor of $\frac 1 2 $  for identical
mesons in the final state is included.}
\label{FlavourFactors}
\end{center}
\end{table}

\begin{table}
\begin{center}
\begin{tabular}{cccccc}
&&&&&\\
Final & Experimental & $^3P_0$ Model &  This & incl. & incl. \\
State&Partial Width&(Refs. \cite{BarnesStrongDecays1,BarnesStrongDecays2})&Work&III&ZRV\\
&&&&&\\
\hline\hline
&&&&&\\
\multicolumn{6}{c}{\fbox{$\rho(770)$\quad,\quad$\Gamma^{\mbox{\scriptsize total}}_{\mbox{\scriptsize exp}}=150.2\pm 0.8$~MeV}}\\
&&&&&\\
$\pi\pi$&$\approx 150$&79&42.9&$\circ$&$\circ$\\
&&&&&\\
\hline
&&&&&\\
\multicolumn{6}{c}{\fbox{$\rho(1450)$\quad,\quad$\Gamma^{\mbox{\scriptsize total}}_{\mbox{\scriptsize exp}}=310\pm 60$~MeV}}\\
&&&&&\\
$\pi\pi$ &seen &74 &2.37 &$\circ$&$\circ$\\
$\omega\pi$ &$<6.2$ &122 &0.00 &$\circ$&$\circ$\\
$\rho\eta$ & $<12.4$&25 & 0.09&$\circ$&$\circ$\\
$KK$ & $<0.50$&35 & 0.03&$\circ$&$\circ$\\
&&&&&\\
\hline
&&&&&\\
\multicolumn{6}{c}{\fbox{$\rho(1700)$\quad,\quad$\Gamma^{\mbox{\scriptsize total}}_{\mbox{\scriptsize exp}}=240\pm 60$~MeV}}\\
&&&&&\\
$\pi\pi$ &seen &48 & 1.16 &$\circ$&$\circ$\\
$\omega\pi$ &seen &35 &1.58 &$\circ$&$\circ$\\
$\rho\eta$ & $<9.6$&16 & 2.82&$\circ$&$\circ$\\
$KK$ & seen&36 & 1.99&$\circ$&$\circ$\\
&&&&&\\
\hline
&&&&&\\
\multicolumn{6}{c}{\fbox{$\rho_3(1690)$\quad,\quad$\Gamma^{\mbox{\scriptsize total}}_{\mbox{\scriptsize exp}}=161\pm 10$~MeV}}\\
&&&&&\\
$\pi\pi$ &$38\pm 3.2$ &59 & 0.82 &$\circ$&$\circ$\\
$\omega\pi$ &$18.3\pm 7.0$ &19 &2.92 &$\circ$&$\circ$\\
$\rho\eta$ & seen&2 & 4.10&$\circ$&$\circ$\\
$KK$ & $2.54¸\pm 0.45$&9 & 6.37&$\circ$&$\circ$\\
$\rho\rho$ & ---& 71 & 55.6&$\circ$&$\circ$\\
\end{tabular}
\caption{Results on strong two--body decays of (excited) isovector $\rho_J$ mesons.
All widths are given in units of [MeV]; the experimental data are
extracted from ref. \cite{PDG2000}. The column ``III'' denotes
whether effects of the instanton--induced three--body interaction are
included ($\bullet$) or not ($\circ$). In the column ``ZRV'', we
indicate whether Zweig rule violation due to the presence of isoscalar
mesons in the instanton--induced mechanism occurs ($\bullet$) or not ($\circ$).}
\label{rho}
\end{center}
\end{table}

\begin{table}
\begin{center}
\begin{tabular}{cccccc}
&&&&&\\
Final & Experimental & $^3P_0$ Model &  This & incl. & incl. \\
State&Partial Width&(Ref. \cite{BarnesStrongDecays2})&Work&III&ZRV\\
&&&&&\\
\hline\hline
&&&&&\\
\multicolumn{6}{c}{\fbox{$\pi(1300)$\quad,\quad$\Gamma^{\mbox{\scriptsize
total}}_{\mbox{\scriptsize exp}}=200\ldots 600$~MeV}}\\
&&&&&\\
$\rho\pi$&seen&209&2.57&$\circ$&$\circ$\\
&&&&&\\
\hline
&&&&&\\
\multicolumn{6}{c}{\fbox{$\pi(1800)$\quad,\quad$\Gamma^{\mbox{\scriptsize total}}_{\mbox{\scriptsize exp}}=210\pm 15$~MeV}}\\
&&&&&\\
$\rho\pi$ &not seen &31 &0.34 &$\circ$&$\circ$\\
$\rho\omega$ & --- &73 &1.46 &$\circ$&$\circ$\\
$f_0(1370)\pi$ &seen &7 &1.12 &$\bullet$&$\bullet$\\
$f_0(1500)\pi$ &seen &--- &3.05 &$\bullet$&$\bullet$\\
$f_2(1270)\pi$ &--- &28 &4.16 &$\circ$&$\circ$\\
$a_0(980)\eta$ &seen &--- &2.79 &$\bullet$&$\bullet$\\
$KK_0^* $ &seen &--- &0.02 &$\bullet$&$\circ$\\
$KK^*$ &not seen &36 &1.76 &$\circ$&$\circ$\\
&&&&&\\
\hline
&&&&&\\
\multicolumn{6}{c}{\fbox{$\pi_2(1670)$\quad,\quad$\Gamma^{\mbox{\scriptsize total}}_{\mbox{\scriptsize exp}}=259\pm 11$~MeV}}\\
&&&&&\\
$\rho\pi$ & 84$\pm$11&118 &12.3 &$\circ$&$\circ$\\
$\rho\omega$ & 7.0$\pm$2.9 &41 &0.11 &$\circ$&$\circ$\\
$\sigma\pi$&34$\pm$16&---&14.7&$\circ$&$\circ$\\
$f_0(1370)\pi$&24$\pm$9&0&0.26&$\circ$&$\circ$\\
$f_2(1270)\pi$&152$\pm$11&75&6.16&$\circ$&$\circ$\\
$KK^*$&11$\pm$4&30&5.18&$\circ$&$\circ$\\
\end{tabular}
\caption{Results on strong two--body decays of (excited) isovector $\pi_J$ mesons.
All widths are given in units of [MeV]; the experimental data are
extracted from ref. \cite{PDG2000}. For further comments, see text and caption
of tab.~\ref{rho}.}
\label{pi}
\end{center}
\end{table}

\begin{table}
\begin{center}
\begin{tabular}{cccccc}
&&&&&\\
Final & Experimental & $^3P_0$ Model &  This & incl. & incl. \\
State&Partial Width&(Refs. \cite{BarnesStrongDecays2,BonnazSilvestreBrac})&Work&III&ZRV\\
&&&&&\\
\hline\hline
&&&&&\\
\multicolumn{6}{c}{\fbox{$\omega(782)$\quad,\quad$\Gamma^{\mbox{\scriptsize
total}}_{\mbox{\scriptsize exp}}=8.44\pm 0.09$~MeV}}\\
&&&&&\\
$\pi\pi$&0.19$\pm$0.03&---&0.00&$\circ$&$\circ$\\
&&&&&\\
\hline
&&&&&\\
\multicolumn{6}{c}{\fbox{$\omega(1420)$\quad,\quad$\Gamma^{\mbox{\scriptsize total}}_{\mbox{\scriptsize exp}}=174\pm 59$~MeV}}\\
&&&&&\\
$\rho\pi$ &dominant &328 &4.73 &$\circ$&$\circ$\\
$\eta\omega$ & --- &12 &2.82 &$\circ$&$\circ$\\
$b_1(1235)\pi$ &--- &1 &0.17 &$\circ$&$\circ$\\
$KK$ &--- &31 &1.99 &$\circ$&$\circ$\\
&&&&&\\
\hline
&&&&&\\
\multicolumn{6}{c}{\fbox{$\omega_3(1670)$\quad,\quad$\Gamma^{\mbox{\scriptsize total}}_{\mbox{\scriptsize exp}}=168\pm 10$~MeV}}\\
&&&&&\\
$\rho\pi$ &seen &50 &8.76 &$\circ$&$\circ$\\
$\eta\omega$ & --- &2 &4.10 &$\circ$&$\circ$\\
$b_1(1235)\pi$ &possibly seen &7 &3.33 &$\circ$&$\circ$\\
$KK$ &--- &8 &6.37 &$\circ$&$\circ$\\
&&&&&\\
\hline
&&&&&\\
\multicolumn{6}{c}{\fbox{$\phi(1020)$\quad,\quad$\Gamma^{\mbox{\scriptsize total}}_{\mbox{\scriptsize exp}}=4.46\pm 0.03$~MeV}}\\
&&&&&\\
$KK$&$2.19\pm 0.04$&4.08&---&$\circ$&$\circ$\\
$\pi\pi$&$(0.33\pm 0.06)\cdot 10^{-3}$&---&0.00&$\circ$&$\circ$\\
$\omega\pi$&$(0.21\pm 0.09)\cdot 10^{-3}$&---&0.00&$\circ$&$\circ$\\
&&&&&\\
\hline
&&&&&\\
\multicolumn{6}{c}{\fbox{$\phi(1680)$\quad,\quad$\Gamma^{\mbox{\scriptsize total}}_{\mbox{\scriptsize exp}}=150\pm 50$~MeV}}\\
&&&&&\\
$KK$  &dominant&---&7.83&$\circ$&$\circ$\\
$KK^*$&seen&---&4.11&$\circ$&$\circ$\\
&&&&&\\
\hline
&&&&&\\
\multicolumn{6}{c}{\fbox{$\phi_3(1850)$\quad,\quad$\Gamma^{\mbox{\scriptsize total}}_{\mbox{\scriptsize exp}}=87{+28\atop -23}$~MeV}}\\
&&&&&\\
$KK$  &seen&---&26.7&$\circ$&$\circ$\\
$KK^*$&seen&---&13.8&$\circ$&$\circ$\\
\end{tabular}
\caption{Results on strong two--body decays of (excited) isoscalar
$\omega_J$ and $\phi_J$ mesons.
All widths are given in units of [MeV]; the experimental data are
extracted from ref. \cite{PDG2000}. Note that the $\omega(1420)$
is considered to be a dominantly $S$--wave state in ref. \cite{BarnesStrongDecays2}; therefore we
identify it with the second $n\bar n$ excitation in the
$\omega/\phi$ system (see \cite{KollRicken,KollPhD}). For the same
reason, we assume that the $\phi(1680)$ is the second $s\bar s$ excitation in the
$\omega/\phi$ system. For further comments, see text and caption
of tab.~\ref{rho}.}
\label{omegaphi}
\end{center}
\end{table}

\begin{table}
\begin{center}
\begin{tabular}{cccccc}
&&&&&\\
Final & Experimental & $^3P_0$ Model &  This & incl. & incl. \\
State&Partial Width&(Refs. \cite{BarnesStrongDecays1,BarnesStrongDecays2})&Work&III&ZRV\\
&&&&&\\
\hline\hline
&&&&&\\
\multicolumn{6}{c}{\fbox{$h_1(1170)$\quad,\quad$\Gamma^{\mbox{\scriptsize
total}}_{\mbox{\scriptsize exp}}=360\pm 40$~MeV}}\\
&&&&&\\
$\rho\pi$&seen&383&50.6&$\circ$&$\circ$\\
&&&&&\\
\hline
&&&&&\\
\multicolumn{6}{c}{\fbox{$h_1(1700)$\quad,\quad$\Gamma^{\mbox{\scriptsize
total}}_{\mbox{\scriptsize exp}}$ unknown}}\\
&&&&&\\
$\rho\pi$&---&173&5.16&$\circ$&$\circ$\\
$\omega\eta$&---&17&0.99&$\circ$&$\circ$\\
$\rho(1465)\pi$&---&31&15.3&$\circ$&$\circ$\\
$b_1(1235)\pi$&---&0&0.00&$\circ$&$\circ$\\
$KK^*$&---&30&0.52&$\circ$&$\circ$\\
&&&&&\\
\hline
&&&&&\\
\multicolumn{6}{c}{\fbox{$h_3(2050)$\quad,\quad$\Gamma^{\mbox{\scriptsize
total}}_{\mbox{\scriptsize exp}}$ unknown}}\\
&&&&&\\
$\rho\pi$&---&115&6.38&$\circ$&$\circ$\\
$\omega\eta$&---&13&5.91&$\circ$&$\circ$\\
$\rho(1465)\pi$&---&1&13.1&$\circ$&$\circ$\\
$b_1(1235)\pi$&---&0&0.00&$\circ$&$\circ$\\
$KK^*$&---&22&2.73&$\circ$&$\circ$\\
&&&&&\\
\hline
&&&&&\\
\multicolumn{6}{c}{\fbox{$b_1(1235)$\quad,\quad$\Gamma^{\mbox{\scriptsize
total}}_{\mbox{\scriptsize exp}}=142\pm 9$~MeV}}\\
&&&&&\\
$\omega\pi$&seen&143&16.9&$\circ$&$\circ$\\
&&&&&\\
\hline
&&&&&\\
\multicolumn{6}{c}{\fbox{$b_1(1700)$\quad,\quad$\Gamma^{\mbox{\scriptsize
total}}_{\mbox{\scriptsize exp}}$ unknown}}\\
&&&&&\\
$\omega\pi$&---&---&1.72&$\circ$&$\circ$\\
$\rho\eta$&---&18&0.99&$\circ$&$\circ$\\
$\rho\rho$&---&60&0.70&$\circ$&$\circ$\\
$a_2(1320)\pi$&---&67&6.51&$\circ$&$\circ$\\
$KK^*$&---&30&0.52&$\circ$&$\circ$\\
&&&&&\\
\hline
&&&&&\\
\multicolumn{6}{c}{\fbox{$b_3(2050)$\quad,\quad$\Gamma^{\mbox{\scriptsize
total}}_{\mbox{\scriptsize exp}}$ unknown}}\\
&&&&&\\
$\omega\pi$&---&37&2.13&$\circ$&$\circ$\\
$\rho\eta$&---&13&5.91&$\circ$&$\circ$\\
$\rho\rho$&---&33&0.02&$\circ$&$\circ$\\
$a_2(1320)\pi$&---&107&3.59&$\circ$&$\circ$\\
$KK^*$&---&22&2.73&$\circ$&$\circ$\\
\end{tabular}
\caption{Results on strong two--body decays of (excited) isoscalar
$h_J$ and isovector $b_J$ mesons.  
All widths are given in units of [MeV]; the experimental data are
extracted from ref. \cite{PDG2000}. The $n\bar n$ states $h_1(1700)$, $h_3(2050)$,
$b_1(1700)$ and $b_3(2050)$ are not listed in ref. \cite{PDG2000} by the
{\sc Particle Data Group}; however, the assumptions of the $^3P_0$
model concerning their masses fit very well to the results of our
quark model (see \cite{KollRicken,KollPhD}). For further comments, see text and caption
of tab.~\ref{rho}.}
\label{hb}
\end{center}
\end{table}

\begin{table}
\begin{center}
\begin{tabular}{cccccc}
&&&&&\\
Final & Experimental & $^3P_0$ Model &  This & incl. & incl. \\
State&Partial Width&(Refs. \cite{BarnesStrongDecays1,BarnesStrongDecays2,BonnazSilvestreBrac})&Work&III&ZRV\\
&&&&&\\
\hline\hline
&&&&&\\
\multicolumn{6}{c}{\fbox{$a_0(980)$\quad,\quad$\Gamma^{\mbox{\scriptsize
total}}_{\mbox{\scriptsize exp}}=50\ldots 100$~MeV}}\\
&&&&&\\
$\eta\pi$&dominant&---&70.2&$\bullet$&$\bullet$\\
$KK$&seen&---&42.9&$\bullet$&$\circ$\\
&&&&&\\
\hline
&&&&&\\
\multicolumn{6}{c}{\fbox{$a_0(1450)$\quad,\quad$\Gamma^{\mbox{\scriptsize
total}}_{\mbox{\scriptsize exp}}=265\pm 13$~MeV}}\\
&&&&&\\
$\eta\pi$&seen&5&31.4&$\bullet$&$\bullet$\\
$\eta '\pi$&seen&5&12.7&$\bullet$&$\bullet$\\
$KK$&seen&0&64.7&$\bullet$&$\circ$\\
&&&&&\\
\hline
&&&&&\\
\multicolumn{6}{c}{\fbox{$a_1(1260)$\quad,\quad$\Gamma^{\mbox{\scriptsize
total}}_{\mbox{\scriptsize exp}}=250\ldots 600$~MeV}}\\
&&&&&\\
$\rho\pi$&seen&545&11.9&$\circ$&$\circ$\\
$\sigma\pi$&seen&---&7.18&$\circ$&$\circ$\\
&&&&&\\
\hline
&&&&&\\
\multicolumn{6}{c}{\fbox{$a_1(1640)$\quad,\quad$\Gamma^{\mbox{\scriptsize
total}}_{\mbox{\scriptsize exp}}=300\pm 50$~MeV}}\\
&&&&&\\
$f_2(1270)\pi$&seen&39&0.32&$\circ$&$\circ$\\
$\sigma\pi$&seen&---&0.58&$\circ$&$\circ$\\
&&&&&\\
\hline
&&&&&\\
\multicolumn{6}{c}{\fbox{$a_2(1320)$\quad,\quad$\Gamma^{\mbox{\scriptsize
total}}_{\mbox{\scriptsize exp}}=104.7\pm 1.9$~MeV}}\\
&&&&&\\
$\rho\pi$&$77.3\pm 5.4$&34.3&15.0&$\circ$&$\circ$\\
$\eta\pi$&$16.0\pm 1.3$&8.01&8.67&$\circ$&$\circ$\\
$\eta '\pi$&$0.59\pm0.10$&0.56&1.07&$\circ$&$\circ$\\
$KK$&5.40$\pm$0.88&5.24&11.7&$\circ$&$\circ$\\
&&&&&\\
\hline
&&&&&\\
\multicolumn{6}{c}{\fbox{$a_4(2040)$\quad,\quad$\Gamma^{\mbox{\scriptsize
total}}_{\mbox{\scriptsize exp}}=361\pm 50$~MeV}}\\
&&&&&\\
$\rho\pi$&seen&33&2.17&$\circ$&$\circ$\\
$\rho_3\pi$&---&2&0.85&$\circ$&$\circ$\\
$f_2(1270)\pi$&seen&10&0.69&$\circ$&$\circ$\\
\end{tabular}
\caption{Results on strong two--body decays of (excited) isovector $a_J$ mesons.  
All widths are given in units of [MeV]; the experimental data are
extracted from ref. \cite{PDG2000}. In the $^3P_0$ model, the second
radial excitation of the $a_0$ meson is assumed to have a mass around $M_{a_0'}\approx
1700$~MeV (see \cite{BarnesStrongDecays2}). Note that the different
versions of the $^3P_0$ model give partly different results for the
decay widths, \eg $\Gamma(a_2\to\rho\pi)=54$~MeV in
ref. \cite{BarnesStrongDecays2} but $\Gamma(a_2\to\rho\pi)=34$~MeV in
ref. \cite{BonnazSilvestreBrac}; in the table, we use the results of
ref. \cite{BonnazSilvestreBrac} concerning the partial widths of the
$a_2(1320)$ meson for comparison. For further comments, see text and caption
of tab.~\ref{rho}.}
\label{a}
\end{center}
\end{table}

\begin{table}
\begin{center}
\begin{tabular}{cccccc}
&&&&&\\
Final & Experimental & $^3P_0$ Model &  This & incl. & incl. \\
State&Partial Width&(Ref. \cite{BarnesStrongDecays1})&Work&III&ZRV\\
&&&&&\\
\hline\hline
&&&&&\\
\multicolumn{6}{c}{\fbox{$f_0(400\ldots 1200)$\quad,\quad$\Gamma^{\mbox{\scriptsize
total}}_{\mbox{\scriptsize exp}}=600\ldots 1000$~MeV}}\\
&&&&&\\
$\pi\pi$&dominant&---&297&$\bullet$&$\bullet$\\
&&&&&\\
\hline
&&&&&\\
\multicolumn{6}{c}{\fbox{$f_0(1370)$\quad,\quad$\Gamma^{\mbox{\scriptsize
total}}_{\mbox{\scriptsize exp}}=275\pm 55$~MeV}}\\
&&&&&\\
$\pi\pi$&       $21.7\pm 9.9$     & 271& 477&$\bullet$&$\bullet$\\  
$\eta\eta$&     $0.41\pm 0.27$    & ---&6.25&$\bullet$&$\bullet$\\  
$\sigma\sigma$& $120.5\pm 45.2$   & ---& ---&$\bullet$&$\bullet$\\  
$\rho\rho$&     $62.2\pm 28.8$    & ---& ---&$\circ$&$\circ$\\      
$KK$&           $5.2\ldots28.4$   & ---&34.7&$\bullet$&$\bullet$\\  
$a_1(1260)\pi$& $14.1\pm 7.2$     & ---&---&$\circ$&$\circ$\\       
&&&&&\\
\hline
&&&&&\\
\multicolumn{6}{c}{\fbox{$f_0(1500)$\quad,\quad$\Gamma^{\mbox{\scriptsize
total}}_{\mbox{\scriptsize exp}}=130\pm 30$~MeV}}\\
&&&&&\\
$\pi\pi$&         $44.1\pm 15.3$    & ---&15.7&$\bullet$&$\bullet$\\
$\pi(1300)\pi$&   $35.5\pm 29.2$ & ---&1.40&$\bullet$&$\bullet$\\
$\eta\eta$&       $3.4\pm 1.2$   & ---&20.2&$\bullet$&$\bullet$\\
$\eta\eta'$&      $2.9\pm 1.0$   & ---&0.08&$\bullet$&$\bullet$\\
$\sigma\sigma$&   $18.6\pm 12.5$ & ---&37.9&$\bullet$&$\bullet$\\
$\rho\rho$&       $8.9\pm 8.2$   & ---& ---&$\circ$&$\circ$\\    
$KK$&             $8.1\pm 2.8$   & ---&27.7&$\bullet$&$\bullet$\\
$a_1(1260)\pi$&   $8.6\pm 6.6$   & ---&2.11&$\circ$&$\circ$\\           
&&&&&\\
\hline
&&&&&\\
\multicolumn{6}{c}{\fbox{$f_1(1285)$\quad,\quad$\Gamma^{\mbox{\scriptsize
total}}_{\mbox{\scriptsize exp}}=24.0\pm 1.2$~MeV}}\\
&&&&&\\
$a_0(980)\pi$&8.64$\pm$1.73&---&0.06&$\circ$&$\circ$\\
&&&&&\\
\hline
&&&&&\\
\multicolumn{6}{c}{\fbox{$f_1(1420)$\quad,\quad$\Gamma^{\mbox{\scriptsize
total}}_{\mbox{\scriptsize exp}}=55.0\pm 2.9$~MeV}}\\
&&&&&\\
$a_0(980)\pi$&---&---&0.00&$\circ$&$\circ$\\
$KK^*$&dominant&---&3.41&$\circ$&$\circ$\\
\end{tabular}
\caption{Results on strong two--body decays of (excited) isoscalar $f_J$ mesons with $J=0,1$.  
All widths are given in units of [MeV]. The experimental data for the
$f_0(1370)$ and $f_0(1500)$ decays are
extracted from ref. \cite{Abele}; the rest is given in
ref. \cite{PDG2000}. Note that the
$f_0(400\ldots1200)$ meson is usually labeled ``$\sigma$'' in this
work. For further comments, see text and caption
of tab.~\ref{rho}.}
\label{f}
\end{center}
\end{table}

\begin{table}
\begin{center}
\begin{tabular}{cccccc}
&&&&&\\
Final & Experimental & $^3P_0$ Model &  This & incl. & incl. \\
State&Partial Width&(Ref. \cite{BonnazSilvestreBrac})&Work&III&ZRV\\
&&&&&\\
\hline\hline
&&&&&\\
\multicolumn{6}{c}{\fbox{$f_2(1270)$\quad,\quad$\Gamma^{\mbox{\scriptsize
total}}_{\mbox{\scriptsize exp}}=185.1{+3.4\atop -2.6}$~MeV}}\\
&&&&&\\
$\pi\pi$   &$156.9{+3.8\atop -1.3}$&144 &6.33&$\circ$&$\circ$\\
$KK$       &$8.6\pm 0.8$           &8.53&11.7&$\circ$&$\circ$\\
$\eta\eta$ &$0.83\pm 0.18$         &1.02&3.48&$\circ$&$\circ$\\
&&&&&\\
\hline
&&&&&\\
\multicolumn{6}{c}{\fbox{$f_2'(1525)$\quad,\quad$\Gamma^{\mbox{\scriptsize
total}}_{\mbox{\scriptsize exp}}=76\pm 10$~MeV}}\\
&&&&&\\
$\pi\pi$   &$0.60\pm 0.12$  & ---&0.00&$\circ$&$\circ$\\
$KK$       &$65{+5\atop -4}$&82.8&60.5&$\circ$&$\circ$\\
$\eta\eta$ &$7.6\pm 2.5$    &7.84&7.86&$\circ$&$\circ$\\
&&&&&\\
\hline
&&&&&\\
\multicolumn{6}{c}{\fbox{$f_4(2050)$\quad,\quad$\Gamma^{\mbox{\scriptsize
total}}_{\mbox{\scriptsize exp}}=222\pm 19$~MeV}}\\
&&&&&\\
$\pi\pi$   &seen& ---&0.29&$\circ$&$\circ$\\
$KK$       &seen& ---&1.86&$\circ$&$\circ$\\
&&&&&\\
\hline
&&&&&\\
\multicolumn{6}{c}{\fbox{$f_4(2220)$\quad,\quad$\Gamma^{\mbox{\scriptsize
total}}_{\mbox{\scriptsize exp}}=23{+8\atop -7}$~MeV}}\\
&&&&&\\
$\pi\pi$   &$37.7\pm 4.6$&53.1&0.00&$\circ$&$\circ$\\
$KK$       &1.51$+0.77\atop -0.42$  &0.25&8.74&$\circ$&$\circ$\\
$\eta\eta$ & 0.47$\pm$ 0.18 &4.62&1.17&$\circ$&$\circ$\\
$\omega\omega$&57.7$\pm 14.2$&23.6&0.00&$\circ$&$\circ$\\
$a_2(1320)\pi$&seen&---&0.00&$\circ$&$\circ$\\
&&&&&\\
\hline
&&&&&\\
\multicolumn{6}{c}{\fbox{$f_6(2510)$\quad,\quad$\Gamma^{\mbox{\scriptsize
total}}_{\mbox{\scriptsize exp}}=255\pm 40$~MeV}}\\
&&&&&\\
$\pi\pi$   &15.3$\pm $3.5& ---&0.02&$\circ$&$\circ$\\
$KK$       &---& ---&0.11&$\circ$&$\circ$\\
\end{tabular}
\caption{Results on strong two--body decays of (excited) isoscalar $f_J$ mesons with $J=2,4,6$.  
All widths are given in units of [MeV]; the experimental data are
extracted from ref. \cite{PDG2000}. According to the {\sc
Particle Data Group}, the total angular momentum of the $f_J(2220)$
could either be $J=2$ or $J=4$ (see \cite{PDG2000}); here, we consider this meson as the
first $s\bar s$ excitation of the $f_4$ system with $I(J^{\pi
c})=0(4^{++})$. For further comments, see text and caption
of tab.~\ref{rho}.}
\label{f2}
\end{center}
\end{table}

\begin{table}
\begin{center}
\begin{tabular}{cccccc}
&&&&&\\
Final & Experimental & $^3P_0$ Model &  This & incl. & incl. \\
State&Partial Width&(Ref. \cite{BonnazSilvestreBrac})&Work&III&ZRV\\
&&&&&\\
\hline\hline
&&&&&\\
\multicolumn{6}{c}{\fbox{$K(1460)$\quad,\quad$\Gamma^{\mbox{\scriptsize
total}}_{\mbox{\scriptsize exp}}\approx 250\ldots 260 $~MeV}}\\
&&&&&\\
$K\rho$&seen&---&56.7&$\circ$&$\circ$\\
$K^*(892)\pi$&seen&---&0.11&$\circ$&$\circ$\\
$K^*_0(1430)\pi$&seen&---&0.01&$\bullet$&$\circ$\\
&&&&&\\
\hline
&&&&&\\
\multicolumn{6}{c}{\fbox{$K_1(1270)$\quad,\quad$\Gamma^{\mbox{\scriptsize
total}}_{\mbox{\scriptsize exp}}=90\pm 20 $~MeV}}\\
&&&&&\\
$K\rho$&37.8$\pm$10.0&118&29.0&$\circ$&$\circ$\\
$K^*(892)\pi$&14.4$\pm$ 5.5&29.6&14.1&$\circ$&$\circ$\\
$K\omega$&9.9$\pm$ 2.8&23.0&7.69&$\circ$&$\circ$\\
&&&&&\\
\hline
&&&&&\\
\multicolumn{6}{c}{\fbox{$K_1(1400)$\quad,\quad$\Gamma^{\mbox{\scriptsize
total}}_{\mbox{\scriptsize exp}}=174\pm 13 $~MeV}}\\
&&&&&\\
$K\rho$&2$\pm$1&77.8&22.0&$\circ$&$\circ$\\
$K^*(892)\pi$&117$\pm$ 10&319&13.9&$\circ$&$\circ$\\
$K\omega$&23$\pm$ 12&31.6&7.40&$\circ$&$\circ$\\
&&&&&\\
\hline
&&&&&\\
\multicolumn{6}{c}{\fbox{$K_2(1580)$\quad,\quad$\Gamma^{\mbox{\scriptsize
total}}_{\mbox{\scriptsize exp}}\approx 110 $~MeV}}\\
&&&&&\\
$K^*(892)\pi$&seen&---&0.69&$\circ$&$\circ$\\
$K^*_2(1430)\pi$&possibly seen&---&0.74&$\circ$&$\circ$\\
$K\phi$&---&---&0.78&$\circ$&$\circ$\\
$K\omega$&---&---&5.48&$\circ$&$\circ$\\
&&&&&\\
\hline
&&&&&\\
\multicolumn{6}{c}{\fbox{$K_2(1770)$\quad,\quad$\Gamma^{\mbox{\scriptsize
total}}_{\mbox{\scriptsize exp}}=186\pm 14 $~MeV}}\\
&&&&&\\
$K^*(892)\pi$&seen&---&6.84&$\circ$&$\circ$\\
$K^*_2(1430)\pi$&dominant&---&2.28&$\circ$&$\circ$\\
$K\phi$&seen&---&       4.37&$\circ$&$\circ$\\
$K\omega$&seen&---&4.91&$\circ$&$\circ$\\
\end{tabular}
\caption{Results on strong two--body decays of (excited) isodublet $K_J$ mesons.  
All widths are given in units of [MeV]; the experimental data are
extracted from ref. \cite{PDG2000}. For
further comments, see text and caption of tab.~\ref{rho}.}
\label{K}
\end{center}
\end{table}

\begin{table}
\begin{center}
\begin{tabular}{cccccc}
&&&&&\\
Final & Experimental & $^3P_0$ Model &  This & incl. & incl. \\
State&Partial Width&(Ref. \cite{BonnazSilvestreBrac})&Work&III&ZRV\\
&&&&&\\
\hline\hline
&&&&&\\
\multicolumn{6}{c}{\fbox{$K_0^*(1430)$\quad,\quad$\Gamma^{\mbox{\scriptsize
total}}_{\mbox{\scriptsize exp}}=294\pm 23$~MeV}}\\
&&&&&\\
$K\pi$&274$\pm$37&455.8&274&$\bullet$&$\circ$\\
&&&&&\\
\hline
&&&&&\\
\multicolumn{6}{c}{\fbox{$K_0^*(1950)$\quad,\quad$\Gamma^{\mbox{\scriptsize
total}}_{\mbox{\scriptsize exp}}=201\pm 86$~MeV}}\\
&&&&&\\
$K\pi$&105$\pm$52&---&80.5&$\bullet$&$\circ$\\
&&&&&\\
\hline
&&&&&\\
\multicolumn{6}{c}{\fbox{$K^*(892)$\quad,\quad$\Gamma^{\mbox{\scriptsize
total}}_{\mbox{\scriptsize exp}}=50.8\pm 0.9$~MeV}}\\
&&&&&\\
$K\pi$&$\approx$50&33.9&48.3&$\circ$&$\circ$\\
&&&&&\\
\hline
&&&&&\\
\multicolumn{6}{c}{\fbox{$K^*(1410)$\quad,\quad$\Gamma^{\mbox{\scriptsize
total}}_{\mbox{\scriptsize exp}}=232\pm 21$~MeV}}\\
&&&&&\\
$K^*(892)\pi$ &$> 93$      &--- &0.00&$\circ$&$\circ$\\
$K\pi$        &15.3$\pm$3.3&23.3&3.51&$\circ$&$\circ$\\
$K\rho$       &$<16.2$     &--- &0.18&$\circ$&$\circ$\\
&&&&&\\
\hline
&&&&&\\
\multicolumn{6}{c}{\fbox{$K^*(1680)$\quad,\quad$\Gamma^{\mbox{\scriptsize
total}}_{\mbox{\scriptsize exp}}=322\pm 110$~MeV}}\\
&&&&&\\
$K^*(892)\pi$ &96.3${+33\atop -36}$ &44.5 &1.27&$\circ$&$\circ$\\
$K\pi$        &124.6$\pm 42$        &107.7&2.71&$\circ$&$\circ$\\
$K\rho$       &101.1${+38\atop -35}$&39.3 &2.21&$\circ$&$\circ$\\
&&&&&\\
\hline
&&&&&\\
\multicolumn{6}{c}{\fbox{$K_2^*(1430)$\quad,\quad$\Gamma^{\mbox{\scriptsize
total}}_{\mbox{\scriptsize exp}}=105\pm 10$~MeV}}\\
&&&&&\\
$K^*(892)\pi$ &$24.3\pm 1.6$           &29.1 &6.75&$\circ$&$\circ$\\
$K\pi$        &$49.1\pm 1.8$           &77.6 &14.1&$\circ$&$\circ$\\
$K\rho$       &$8.5\pm 0.8$            &17.6 &3.89&$\circ$&$\circ$\\
$K\omega$     &$2.9\pm 0.8$            &2.7  &1.30&$\circ$&$\circ$\\
$K\eta$       &$0.15{+0.33\atop -0.10}$&3.4  &0.06&$\circ$&$\circ$\\
&&&&&\\
\hline
&&&&&\\
\multicolumn{6}{c}{\fbox{$K_2^*(1980)$\quad,\quad$\Gamma^{\mbox{\scriptsize
total}}_{\mbox{\scriptsize exp}}=373\pm 70$~MeV}}\\
&&&&&\\
$K^*(892)\pi$ &seen&--- &0.00&$\circ$&$\circ$\\
$K\pi$        &---&---&1.15&$\circ$&$\circ$\\
$K\rho$       &seen&--- &0.53&$\circ$&$\circ$\\
\end{tabular}
\caption{Results on strong two--body decays of (excited) isodublet $K^*_J$ mesons with $J=0,1,2$.  
All widths are given in units of [MeV]; the experimental data are
extracted from ref. \cite{PDG2000}. For
further comments, see text and caption of tab.~\ref{rho}.}
\label{Kstar}
\end{center}
\end{table}

\begin{table}
\begin{center}
\begin{tabular}{cccccc}
&&&&&\\
Final & Experimental & $^3P_0$ Model &  This & incl. & incl. \\
State&Partial Width&(Ref. \cite{BonnazSilvestreBrac})&Work&III&ZRV\\
&&&&&\\
\hline\hline
&&&&&\\
\multicolumn{6}{c}{\fbox{$K_3^*(1780)$\quad,\quad$\Gamma^{\mbox{\scriptsize
total}}_{\mbox{\scriptsize exp}}=159\pm 21$~MeV}}\\
&&&&&\\
$K\rho$           &49.3$\pm$15.7 &30.0&9.17&$\circ$&$\circ$\\
$K^*(892)\pi$     &31.8$\pm$9.0  &37.1&2.84&$\circ$&$\circ$\\
$K\pi$            &29.9$\pm$4.3  &46.9&3.83&$\circ$&$\circ$\\
$K\eta$           &47.7$\pm$21.6 &8.4 &12.8&$\circ$&$\circ$\\
$K^*_2(1430)\pi$  &$< 25.4$      &--- &0.14&$\circ$&$\circ$\\
$K^*(892)\rho$    &---  &---&44.3&$\circ$&$\circ$\\
$K^*(892)\omega$    &---  &---&14.8&$\circ$&$\circ$\\
&&&&&\\
\hline
&&&&&\\
\multicolumn{6}{c}{\fbox{$K_4^*(2045)$\quad,\quad$\Gamma^{\mbox{\scriptsize
total}}_{\mbox{\scriptsize exp}}=198\pm 30$~MeV}}\\
&&&&&\\
$K\pi$            &19.6$\pm$3.8  &21.7&1.10&$\circ$&$\circ$\\
$K^*(892)\phi$    &2.8$\pm$1.4  &---&1.13&$\circ$&$\circ$\\
&&&&&\\
\hline
&&&&&\\
\multicolumn{6}{c}{\fbox{$K_5^*(2380)$\quad,\quad$\Gamma^{\mbox{\scriptsize
total}}_{\mbox{\scriptsize exp}}=178\pm 49$~MeV}}\\
&&&&&\\
$K\pi$            &10.9$\pm$3.7  &---&0.33&$\circ$&$\circ$\\
\end{tabular}
\caption{Results on strong two--body decays of (excited) isodublet $K^*_J$ mesons with $J=3,4,5$.  
All widths are given in units of [MeV]; the experimental data are
extracted from ref. \cite{PDG2000}. For
further comments, see text and caption of tab.~\ref{rho}.}
\label{Kstar2}
\end{center}
\end{table}

\begin{table}
\begin{center}
\begin{tabular}{cccccc}
&&&&&\\
Final & Experimental & $^3P_0$ Model &  This & incl. & incl. \\
State&Partial Width&(Ref. \cite{BarnesStrongDecays2})&Work&III&ZRV\\
&&&&&\\
\hline\hline
&&&&&\\
\multicolumn{6}{c}{\fbox{$\eta '(958)$\quad,\quad$\Gamma^{\mbox{\scriptsize
total}}_{\mbox{\scriptsize exp}}=0.20\pm 0.02 $~MeV}}\\
&&&&&\\
$\rho\pi$&$< 0.008$&---&0.00&$\circ$&$\circ$\\
&&&&&\\
\hline
&&&&&\\
\multicolumn{6}{c}{\fbox{$\eta (1440)$\quad,\quad$\Gamma^{\mbox{\scriptsize
total}}_{\mbox{\scriptsize exp}}=50\ldots 80 $~MeV}}\\
&&&&&\\
$KK^*$&$21.6\pm 5.2$&---&21.8&$\circ$&$\circ$\\
$a_0(980)\pi$&$26.6\pm 7.0$&---&18.8&$\bullet$&$\bullet$\\
$\sigma\eta$&$27.0\pm 6.0$&---&43.9&$\bullet$&$\bullet$\\
&&&&&\\
\hline
&&&&&\\
\multicolumn{6}{c}{\fbox{$\eta _2(1645)$\quad,\quad$\Gamma^{\mbox{\scriptsize
total}}_{\mbox{\scriptsize exp}}=180{+22\atop -20}$~MeV}}\\
&&&&&\\
$a_0(980)\pi$&seen&---&20.8&$\circ$&$\circ$\\
$a_2(1320)\pi$&seen&189&18.5&$\circ$&$\circ$\\
$KK^*$&seen&26&5.18&$\circ$&$\circ$\\
$\rho\rho$&---&33&0.16&$\circ$&$\circ$\\
&&&&&\\
\hline
&&&&&\\
\end{tabular}
\caption{Results on strong two--body decays of (excited) isoscalar $\eta_J$ mesons.  
All widths are given in units of [MeV]. The
experimental data for the $\eta(1440)$ partial widths are extracted
from ref. \cite{Suh}; the rest is given in ref. \cite{PDG2000}. For
further comments, see text and caption of tab.~\ref{rho}.}
\label{eta}
\end{center}
\end{table}

\cleardoublepage

\subsection{Figures}

\end{appendix}

\end{document}